\documentclass[twocolumn]{aastex61}

\usepackage{graphicx}				
\usepackage{xcolor}
\usepackage[sort&compress]{natbib}
\usepackage[hang,flushmargin]{footmisc}
\usepackage[counterclockwise]{rotating}
\usepackage{amsmath,natbib,graphicx}
\usepackage{lipsum}
\usepackage{gensymb}

\shortauthors{Checlair, et al.}
\shorttitle{Frequency of Earth-like planets with LUVOIR vs. HabEx}
\usepackage{enumitem,xcolor}
\begin{document}
\graphicspath{ {./} }
\DeclareGraphicsExtensions{.pdf,.eps,.png}

\title{Probing the capability of future direct imaging missions to spectrally constrain the frequency of Earth-like planets}
\accepted{at The Astronomical Journal, January 7 2021}

\author{Jade H. Checlair}
\affiliation{Department of the Geophysical Sciences, University of
  Chicago, Chicago, IL 60637}

\author{Geronimo L. Villanueva}
\affiliation{NASA Goddard Space Flight Center, Greenbelt, MD 20771} 

\author{Benjamin P.C. Hayworth}
\affiliation{Department of Geosciences, Pennsylvania State University, University Park, PA 16802} 

\author{Stephanie L. Olson}
\affiliation{Department of Earth, Atmospheric, and Planetary Sciences, Purdue University, West Lafayette, IN 47907}
\affiliation{Department of the Geophysical Sciences, University of
  Chicago, Chicago, IL 60637} 

\author{Thaddeus D. Komacek}
\affiliation{Department of the Geophysical Sciences, University of
  Chicago, Chicago, IL 60637} 
  
 \author{Tyler D. Robinson}
\affiliation{Department of the Astronomy and Planetary Science, Northern Arizona University, Flagstaff, AZ 86011}  

\author{Predrag Popovic}
\affiliation{Institut de Physique du Globe de Paris, Paris, France}
\affiliation{Department of the Geophysical Sciences, University of
  Chicago, Chicago, IL 60637} 

\author{Huanzhou Yang}
\affiliation{Department of the Geophysical Sciences, University of
  Chicago, Chicago, IL 60637} 

\author{Dorian S. Abbot}
\affiliation{Department of the Geophysical Sciences, University of
  Chicago, Chicago, IL 60637}

\correspondingauthor{Jade H. Checlair}
\email{jadecheclair@uchicago.edu}

\begin{abstract}
A critical question in astrobiology is whether exoEarth candidates (EECs) are Earth-like, in that they originate life that progressively oxygenates their atmospheres similarly to Earth. We propose answering this question statistically by searching for O$_2$ and O$_3$ on EECs with missions such as HabEx or LUVOIR. We explore the ability of these missions to constrain the fraction, f$_E$, of EECs that are Earth-like in the event of a null detection of O$_2$ or O$_3$ on all observed EECs. We use the Planetary Spectrum Generator to simulate observations of EECs with O$_2$ and O$_3$ levels based on Earth's history. We consider four instrument designs: LUVOIR-A (15m), LUVOIR-B (8m), HabEx with a starshade (4m, ``HabEx/SS''), HabEx without a starshade (4m, ``HabEx/no-SS’’); as well as three estimates of the occurrence rate of EECs ($\eta_{earth}$): 24\%, 5\%, and 0.5\%. In the case of a null-detection, we find that for $\eta_{earth}$ = 24\%, LUVOIR-A, LUVOIR-B, and HabEx/SS would constrain f$_E$ to $\leq$ 0.094, $\leq$ 0.18, and $\leq$ 0.56, respectively. This also indicates that if f$_E$ is greater than these upper limits, we are likely to detect O$_3$ on at least 1 EEC. Conversely, we find that HabEx/no-SS cannot constrain f$_E$, due to the lack of an coronagraph ultraviolet channel. For $\eta_{earth}$ = 5\%, only LUVOIR-A and LUVOIR-B would be able to constrain f$_E$, to $\leq$ 0.45 and $\leq$ 0.85, respectively. For $\eta_{earth}$ = 0.5\%, none of the missions would allow us to constrain f$_E$, due to the low number of detectable EECs. We conclude that the ability to constrain f$_E$ is more robust to uncertainties in $\eta_{earth}$ for missions with larger aperture mirrors. However all missions are susceptible to an inconclusive null detection if $\eta_{earth}$ is sufficiently low.

\end{abstract}

\section{Introduction}
\label{sec:intro}

The past decade has been incredibly productive in exoplanet research, with more than 2500 exoplanets confirmed by NASA’s Kepler and K2 missions. Of these, at least 30 planets with radii less than twice Earth's radius were found orbiting in the habitable zone of their star \citep[e.g.,][]{burke2015terrestrial, dressing2015occurrence}. More recently, a number of Earth-sized exoplanets have been found orbiting in the habitable zone of nearby stars \citep[e.g.,][]{gillon2017seven, dittmann2017temperate}. These are prime targets for future instruments to study in more detail and determine whether they are truly habitable, or even inhabited \citep[e.g.,][]{kreidberg2016prospects, meadows2018habitability}. NASA’s upcoming James Webb Space Telescope should dramatically increase our ability to find and characterize terrestrial exoplanets, but all will be orbiting M-stars and they will mostly be too hot for habitability \citep[e.g.,][]{deming2009discovery, cowan2015characterizing}. The next generation of proposed instruments for the NASA Astronomy and Astrophysics Decadal Survey includes two space-based telescopes: HabEx \citep{gaudi2020habitable}, and LUVOIR \citep{luvoir2019luvoir}. These instruments will allow us to characterize the atmospheres of habitable zone planets orbiting Sun-like stars via direct imaging. 

Life can have a measurable impact on the composition of its host planet's atmosphere. A long-standing goal in astrobiology is to spectrally determine the presence of life via biosignatures in the atmosphere \citep{schwieterman2018exoplanet}. 
Several potential biosignatures have been proposed, such as detecting trace amounts of biologically derived molecules \citep[e.g.,][]{seager2005vegetation,meadows2008planetary,seager2010exoplanet,seager2016toward,sousa2019molecular}, measuring thermodynamic chemical disequilibrium between atmospheric species \citep{lovelock1965physical, krissansen2018disequilibrium}, observing a ``red-edge'' in the atmospheric spectrum \citep{seager2005vegetation}, and detecting seasonal variation \citep{olson2018atmospheric}. Here, we focus on perhaps the most robust biosignatures for G-star planets, O$_2$ and O$_3$ \citep[e.g.,][]{owen1980search,sagan1993search,des2002remote,meadows2008planetary,meadows2017reflections,schwieterman2018exoplanet}. We find that O$_3$ is always easier to detect than O$_2$ at UV-VIS wavelengths, so we expect that O$_3$ is the main signal for an oxygenated atmosphere that will be used in future space telescope missions.

Though extensive and ongoing work is being done to determine abiotic sources of O$_2$ \citep{harman2015abiotic, tian2014high, meadows2017reflections}, most potential ``false-positive'' scenarios explored in the literature so far occur for either M-dwarf planets or planets that have since gone through a moist or runaway greenhouse phase, rendering them outside the classical habitable zone. False positive scenarios for Sun-like star planets with an Earth-like inventory of non-condensing gases have not yet been identified \citep{wordsworth2014abiotic,meadows2018exoplanet,harman2018abiotic}, therefore, we expect that we will be able to interpret O$_2$ and O$_3$ detections confidently within their broader chemical and planetary context. Regardless, false positives are not considered in this work as our primary concern is to determine what conclusions could be drawn from a potential null detection of O$_2$ and/or O$_3$ on all of the detected planets.

As future instruments are being launched and developed, the consensus is that planets that are Earth-sized, terrestrial, and orbiting in the habitable zone of their stars are great targets for the search for life. We will refer to Earth-sized habitable zone planets as ``exoEarth candidates'' (EECs), although this name does not indicate that they are ``Earth-like.'' We will call an ``Earth-like planet'' an EEC that develops Earth-like O$_2$-producing life that oxygenates its atmosphere roughly following Earth's oxygenation history. Following this statistical definition, Earth-like EECs would start at negligible levels of atmospheric O$_2$ such as Earth did during its Hadean and Archean eras, and progressively develop an oxygenated atmosphere. A number of Earth-like EECs would be expected to be found in a Hadean or Archean-like era and so to lack remotely detectable levels of O$_2$ and O$_3$ as Earth did during its Hadean and Archean eras. This definition of Earth-like EECs is therefore not meant to be used for individual planets, but instead is purely statistical. If we do not detect O$_2$ or O$_3$ on an individual planet, we cannot know whether it resembles Hadean or Archean Earth and will eventually develop an oxygenated atmosphere, or if it is sterile and never will. Instead we'll simply know that at this stage in its history it does not currently have remotely detectable levels of oxygen in its atmosphere. 

We will refer to the fraction of EECs that are Earth-like as f$_E$. If EECs are generally unlikely to be Earth-like (low f$_E$), it could either mean that they usually do not originate Earth-like life in the first place, or that although they do originate life, O$_2$ levels tend to never increase past Archean-like levels (either because oxygenic photosynthesis is rare or because oxygenic photosynthesis does not always manifest as planetary oxygenation). In both scenarios, EECs would not statistically be considered Earth-like, and we will not detect O$_2$ or O$_3$. The question we are trying to answer is: \textit{Could future direct imaging instruments, such as HabEx and LUVOIR, constrain the frequency of Earth-like EECs, even if they do not detect any?} 

A critical consideration in this work is the possibility of mission-level false negatives: cases where it is common for EECs to be Earth-like but we do not detect O$_2$ or O$_3$ on any of them. We will be adopting a statistical approach to address this problem \citep{bean2017statistical, checlair2019statistical, bixel2020testing}. This means that we will not investigate false negatives on particular planets. Instead, we will try to determine whether there might be mission-level false negatives for a particular mission based on all of the information that we can gain from all of the EECs that this mission can be expected to observe. Even with a large sample of EECs, we may not detect O$_2$ or O$_3$ with either LUVOIR or HabEx if the origination of life is uncommon or if EECs do not generally develop oxygenic photosynthesis. The frequency of life origination on habitable planets is highly uncertain \citep[e.g.,][]{sandberg2018dissolving}, so this is a scenario that should be seriously considered. Because it is possible that Earth-like EECs are very rare, we should design an instrument that would discern this, rather than possibly being a mission-level false negative scenario. 

Using a statistical methodology with future direct imaging instruments will maximize the scientific return of these missions by allowing us to test theories of planetary habitability. This article focuses on statistically testing whether EECs are generally Earth-like, which necessitates a large enough sample of EECs so that mission-level false negative scenarios are unlikely and that we may constrain f$_E$. In previous work, \citet{bixel2020testing} proposed statistically testing the ``age-oxygen'' correlation with future observatories to determine whether we could place constraints on the amount of O$_2$ on EECs given the system age. This would then allow us to prioritize systems of a certain age for characterization when a large sample of EECs is available. \citet{bixel2020testing} found that testing this hypothesis would require a large number of EECs ($\sim$200 EECs if 10\% of detectable EECs have detectable O$_2$ or O$_3$, $\sim$20-40 EECs if 50\% of detectable EECs have detectable O$_2$ or O$_3$) and may be therefore only be possible with a LUVOIR-A-like instrument, which is expected to detect more EECs, assuming most EECs have detectable O$_2$ or O$_3$.

The occurrence rate of EECs ($\eta_{\earth}$) is difficult to constrain as there have not yet been any Earth-sized exoplanets detected in the habitable zone of G-stars. Estimating it therefore requires an extrapolation from Kepler data based on the population of small short-period planets. Based on this sample, many different estimates have been published in the literature that vary significantly \citep[e.g.,][]{catanzarite2011occurrence,petigura2013prevalence,burke2015terrestrial,mulders2018exoplanet}. To come to a community consensus, the NASA funded Exoplanet Exploration Program Analysis Group (ExoPAG) led Study Analysis Group 13 (SAG13) to compile published occurence rates from the literature and proposed average value of $\eta_{\earth}$ with uncertainties \citep{belikov2017exoplanet,kopparapu2018exoplanet}. Based on the SAG13 study, \citet{stark2019exoearth} adopted boundaries for planet radii of $8((a/1 AU))^{-0.5}$ R$_{\earth}$ $\leq$ R $\leq$ 1.4 R$_{\earth}$, and for semi-major axes of 0.95-1.67 AU, and integrated the SAG13 occurence rates over these boundaries to estimate $\eta_{\earth}$ $\sim$ 24$\pm^{46}_{16}$\%. However, recent studies from \citet{pascucci2019impact} and \citet{neil2020joint} showed that extrapolating from that Kepler sample is problematic as it is contaminated by stripped sub-Neptune cores and this sample can therefore not be reliably used to estimate $\eta_{\earth}$. This suggests that the estimate used by \citet{stark2019exoearth} may be overly optimistic. \citet{pascucci2019impact} re-evaluated $\eta_{\earth}$ using exoplanets at larger separations and excluding short-period planets, which may be contaminated by stripped cores, and estimated $\eta_{\earth}$ to be $\sim 5-10$\%. More recently, \citet{neil2020joint} used Bayesian models fit to the Kepler data to calculate occurrence rates of planets in different regimes and found that using models with envelope mass loss predicts an order of magnitude drop in $\eta_{\earth}$, down to $\sim0.5$\%. In this work, we consider three estimates of $\eta_{\earth}$: 24\% \citep{stark2019exoearth}, 5\% \citep{pascucci2019impact}, and 0.5\% \citep{neil2020joint}; and we explore how these estimates affect the ability of future direct imaging missions to constrain the fraction of Earth-like EECs in the case of a null detection of O$_2$ and O$_3$ on all of the observed EECs. 

This paper is organized as follows. In Section~\ref{sec:methods} we outline how we determine the number of EECs on which LUVOIR and HabEx could detect O$_2$ and/or O$_3$ and the constraints that could be placed on f$_E$ in the case of a null detection. In Section~\ref{sec:results} we first theoretically solve for the probability of a mission-level false negative scenario for any mean EEC yield. We then present the integration times necessary for LUVOIR and HabEx to detect O$_2$ and O$_3$ at 5-$\sigma$. Next, we present the number of EECs on which LUVOIR and HabEx could detect O$_2$ and/or O$_3$ for different values of $\eta_{\earth}$. We end that section by discussing how we could use these observations to constrain the fraction, f$_E$, of EECs that are Earth-like and how this may be affected by the adopted estimate of $\eta_{\earth}$. In Section~\ref{sec:discussion} we discuss some implications and caveats of our work, and we summarize results in Section~\ref{sec:conclusions}. 

\section{Methods}
\label{sec:methods}

\subsection{Overview}
\label{sec:overview}
 
 \begin{figure*}
\begin{center}
  \includegraphics[width=1.\textwidth]{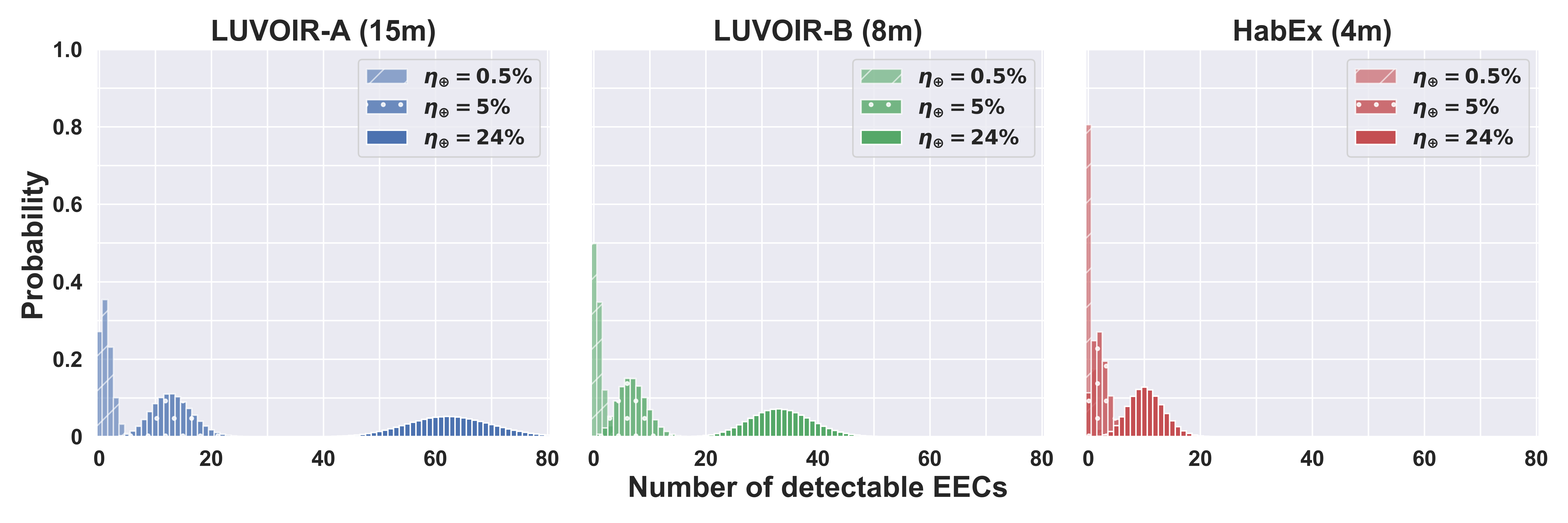}
 \end{center}
  \caption{\textbf{Probability distributions of the number of EECs that would be detected by 15 m segmented on-axis LUVOIR-A (left), 8 m segmented off-axis LUVOIR-B (center), and 4 m monolith off-axis HabEx (right). For $\eta_{\earth}$ = 5\%, there is an 11\% chance of not detecting a single EEC with HabEx. For $\eta_{\earth}$ = 0.5\%, there is a 27\%, 50\%, and 81\% chance of not detecting any EECs with LUVOIR-A, LUVOIR-B, and HabEx, respectively.} All distributions assume a high instrumental throughput. Histograms are based on data from \citet{stark2019exoearth}. Solid histograms assume $\eta_{\earth} = 24$\%, following \citet{stark2019exoearth}, dotted histograms assume $\eta_{\earth} = 5$\%, following \citet{pascucci2019impact}, and barred histograms assume $\eta_{\earth} = 0.5$\%, following \citet{neil2020joint}.}
  \label{fig:Np}
\end{figure*}

To estimate the distribution of the number of detected Earth-like EECs, we perform Monte Carlo simulations where we consider an ensemble of many repeated HabEx and LUVOIR experiments. For each Monte Carlo realization, we draw the number of planets, N$_p$, detected by each instrument and we assign to each detected planet a distance from Earth and an age that are used to determine whether O$_2$ and O$_3$ are detectable on the planet. Below, we first explain how we draw N$_p$ and then we explain how we draw age and distance.

We consider four different mission designs: 15 m segmented on-axis LUVOIR-A, 8 m segmented off-axis LUVOIR-B, 4 m monolith off-axis HabEx with a starshade (``HabEx/SS'', where ``SS'' refers to HabEx's starshade, the complete proposed HabEx mission) and 4 m monolith off-axis HabEx without a starshade (``HabEx/no-SS'', a proposed descoped version of HabEx). We choose these designs because they bracket the range of reasonably likely space-based direct imaging missions over the next few decades. We use the results of \citet{stark2019exoearth} for the expected number of detectable EECs, N$_p$, for each mission design. These expected yields assume $\eta_{\earth}$ = 24\%. To consider lower estimates of $\eta_{\earth}$, we perform Monte Carlo simulations by resampling the yield estimate data from \citet{stark2019exoearth} and weighing each EEC draw by a factor of (5/24) for $\eta_{\earth}$ = 5\% \citep{pascucci2019impact}, and by a factor of (0.5/24) for $\eta_{\earth}$ = 0.5\% \citep{neil2020joint}. The resulting distributions of $N_p$ for LUVOIR-A, LUVOIR-B, and HabEx, for different values of $\eta_{\earth}$, are shown in Figure~\ref{fig:Np}. We note here that we draw values for the number of detectable EECs, N$_p$, for both HabEx/SS and HabEx/no-SS from the same HabEx distribution in Figure~\ref{fig:Np}'s left panel. We set the exposure time to 1000 hours per planet, which is a realistic upper limit for observatories such as LUVOIR and HabEx. Observations may require a shorter integration time, while a longer one, although possible in principle, would likely hamper other mission science objectives. 

For each Monte Carlo realization, we first draw a value for $N_p$ from Figure~\ref{fig:Np} for each instrument. This is the number of planets that will be detected using either instrument for each Monte Carlo realization. We then draw an age for each planet from a uniform distribution between 4.54 Gyr old and 0 Gyr old, to cover Earth's history. The age we draw belongs to one of Earth's eras: Hadean (4.54-4 Gya, 11.9\% probability of being drawn), Archean (4-2.5 Gya, 33.0\% probability of being drawn), pre-GOE Proterozoic (2.5-2.3 Gya, 4.4\% probability of being drawn), post-GOE Proterozoic (2.3-0.5 Gya, 39.7\% probability of being drawn), Phanerozoic (0.5-0 Gya, 11.0\% probability of being drawn). We then draw a distance in parsecs for each of these $N_p$ planets. We draw these distances from a list of targets for each mission provided by Chris Stark \citep{stark2019exoearth}, weighted by the habitable zone yield estimates, $\eta_{\earth}$, for each target. Considering the planet's era and distance, we calculate the Signal-to-Noise Ratio (SNR) for O$_2$ and O$_3$ detections with each telescope design given an exposure time of 1000 hours. We define a detectability SNR threshold of 5.0 for O$_2$ and/or O$_3$ to be considered detectable, and we count the number of EECs on which this condition is met. We find that the SNR is always higher for O$_3$ for all the mission designs considered. We repeat this for $10^7$ Monte Carlo realizations. 

An important assumption we make is that EECs that are Earth-like develop life and atmospheric O$_2$ following the same trajectory as Earth did: Hadean, Archean, Proterozoic, Phanerozoic. This means that we assume f$_E = 1$ to determine whether there are any mission-level false negative scenarios under this assumption. We view this as a starting assumption that is necessary to make progress, rather than the most likely scenario. We also consider a wide range of Proterozoic O$_2$ levels, which allows us to investigate a large range of O$_2$ history scenarios. We assume that O$_2$ and O$_3$ are undetectable at Hadean, Archean, and pre-GOE Proterozoic levels. We also assume EECs remain inhabited for the current inhabited history of the Earth of $\sim$3.8 Gyr \citep{schidlowski19883, dodd2017evidence}. This of course would depend on the planet's position in its star's habitable zone, as this will determine the length of time the planet remains habitable \citep{kopparapu2013habitable}. Earth will only remain in its own habitable zone for $\sim$1.75 Gyrs before entering a runaway greenhouse climate \citep{rushby2013habitable}. During this time, Earth's atmospheric CO$_2$ should decrease to very low values as a result of the silicate-weathering feedback so that oxygenic photosynthesis by land plants will eventually fail \citep{caldeira1992life}. This would likely result in a major decrease in Earth's atmospheric O$_2$. Since the trajectory of Earth's future atmospheric O$_2$ levels is highly uncertain, here we simply draw O$_2$ levels from Earth's history but recognize this may be an optimistic assumption. 
 
\subsection{Planetary Atmosphere simulations}
\label{sec:atmossim}

We start by generating atmospheric profiles appropriate for Earth throughout its history. We assume a cloud-free atmosphere for all the profiles. We do not generate profiles for the Hadean and Archean, as O$_2$ and O$_3$ concentrations remain below $\sim$10$^{-5}$ Present Atmospheric Level (PAL) for the Archean \citep{kasting1979oxygen, pavlov2002mass}. For the Phanerozoic, we use empirical atmospheric profiles for the Modern Earth provided by NASA's MERRA-2 dataset \citep{gelaro2017Modern, villanueva2018planetary}. For the Proterozoic, we calculate the mixing ratio profiles using a one-dimensional, horizontally averaged photochemical model \citep{segura2007abiotic}. The model has 35 long-lived chemical species, 16 short-lived chemical species, and 220 reactions. A two-stream approximation is used for radiative transfer, using a fixed zenith angle of 50 degrees. The model solves for the steady-state solution at each altitude layer, accounting for chemical reactions, photolytic reactions, and vertical transport parameterized using Earth-like eddy diffusion profiles \citep{segura2007abiotic, harman2015abiotic}. 

\begin{table*}
  \caption{\textbf{Instrument parameters we used in PSG to simulate observations with LUVOIR-A, LUVOIR-B, HabEx/SS, and HabEx/no-SS. The throughput values for $T_{coronagraph}$ and $T_{opt}$ are representative average values over the wavelength ranges. We include details about all assumed throughputs in the Appendix.}}
\label{tab:params}
\centering
\begin{tabular}{l l l l l l}
 Parameter & LUVOIR-A & LUVOIR-B & HabEx/SS & HabEx/no-SS \\
\hline
Diameter  &  15 m &  8 m & 4 m & 4 m\\ 
Wavelength Range   & UV: 0.2 $-$ 0.515 $\micron$  & UV: 0.2 $-$ 0.515 $\micron$ & UV: 0.2 $-$ 0.45 $\micron$ & UV: 0.35 $-$ 0.45 $\micron$ \\
  & VIS: 0.515 $-$ 1.0 $\micron$ & VIS:  0.515 $-$ 1.0 $\micron$ & VIS: 0.45 $-$ 0.975 $\micron$ & VIS: 0.45 $-$ 0.975 $\micron$  \\
  & NIR: 1.0 $-$ 2.0 $\micron$  & NIR: 1.0 $-$ 2.0 $\micron$ & NIR: 0.975 $-$ 1.8 $\micron$ & NIR: 0.975 $-$ 1.8 $\micron$ \\ 
Resolution  & UV: 7 & UV: 7 & UV: 7 & UV: 7 \\
  & VIS: 140 & VIS: 140 & VIS: 140 & VIS: 140 \\
  & NIR: 70 & NIR: 70 & NIR: 40 & NIR: 40\\ 
Exozodi Level &  4.5 & 4.5 & 4.5 & 4.5\\ 
(relative to Solar System) & & & & \\
Contrast  & 1$\times$10$^{-10}$ & 1$\times$10$^{-10}$ & 1$\times$10$^{-10}$ & 2.5$\times$10$^{-10}$\\ 
IWA [$\lambda$/D]  & 4 &  3.5 & 39 (UV), 58 (VIS), 104 (NIR) mas & 2.5\\ 
Read noise [e-]  &  UV: 0 &  UV: 0 &  UV: 0.008 &  UV: 0.008 \\ 
   & VIS: 0 & VIS: 0 & VIS: 0.008 & VIS: 0.008 \\
   & NIR: 2.5 & NIR: 2.5 & NIR: 0.32 & NIR: 0.32 \\
Dark noise [e-/s] &  UV: 3e-5 &  UV: 3e-5 &  UV: 3e-5 &  UV: 3e-5 \\ 
   & VIS: 3e-5 & VIS: 3e-5 & VIS: 3e-5 & VIS: 3e-5 \\
   & NIR: 0.002 & NIR: 0.002 & NIR: 0.005 & NIR: 0.005 \\
$T_{coronagraph}$ & 0.27 & 0.46 & 0.7 & 0.55\\
$T_{opt}$ & UV: 0.13 & UV: 0.13 & UV: 0.38 & UV: 0.09\\
 & VIS: 0.21 & VIS: 0.21 & VIS: 0.27 & VIS: 0.15\\
  & NIR: 0.3 & NIR: 0.3 & NIR: 0.36 & NIR: 0.15\\
\end{tabular}
\end{table*}

Proterozoic O$_2$ levels are poorly constrained. We repeat our calculations for several Proterozoic O$_2$ scenarios ranging from 10$^{-5}$ to 10$^{-1}$ PAL to survey the full range of estimates existing in the literature \citep{pavlov2002mass, planavsky2014low, lyons2014rise, reinhard2017false, olson2018earth}. Here we note that the lower end of that range (10$^{-5}$ PAL) is difficult to explain in biogeochemical and photochemical models if oxygenic photosynthesis was occurring at or near modern rates \citep{ozaki2019sluggish}. The most likely range of Proterozoic O$_2$ may therefore be 10$^{-4}$$-$10$^{-1}$ PAL. We discuss the constraints on Proterozoic O$_2$ further in Section~\ref{sec:discussion}. \citet{reinhard2017false} and \citet{olson2018earth} also surveyed the existing literature for CO$_2$ and CH$_4$ estimates throughout Earth's history, and placed upper and lower bounds on their abundances during each era. They argued for stricter constraints on each of these species' mixing ratios, providing ``preferred ranges'' for the Proterozoic. We use the mid-point values of those ``preferred ranges'' for Proterozoic CO$_2$ ($\sim$2000 $\mu$bar) and CH$_4$ ($\sim$5 $\mu$bar). For our Proterozoic water vapor profile, we assume a moist adiabat with a fixed relative humidity of 0.8. We then calculate the O$_3$ profile based on our background atmosphere (assumed to be 1 bar with N$_2$ as the major constituent). We then calculate the temperature profile using CLIMA, a one-dimensional radiative convective climate model \citep{kopparapu2013habitable}. As the Proterozoic has less O$_3$ than Modern Earth, there is a smaller temperature inversion in the stratosphere. We set the surface albedo to 0.3 for all of our atmospheric profiles.

\subsection{Simulated Observations}
\label{sec:simobs}

We use the Planetary Spectrum Generator (PSG) (\citet{villanueva2018planetary}, https://psg.gsfc.nasa.gov) to simulate observations with 15 m segmented on-axis LUVOIR-A, 8 m segmented off-axis LUVOIR-B,  4 m monolith off-axis HabEx with a starshade (``HabEx/SS'') and 4 m monolith off-axis HabEx without a starshade (``HabEx/no-SS''). For all instruments we set the exposure time to 1000 hours. We choose the instrument parameters based on their reported values in their respective Final Reports \citep{luvoir2019luvoir, gaudi2020habitable}. We summarize these parameters in Table~\ref{tab:params}. 

\subsection{Signal-to-Noise Ratio (SNR) Calculations}
\label{sec:snrcalc}
 
To calculate the SNRs, we simulate two spectra: one absorbing spectrum with all atmospheric species, and one continuum spectrum with all atmospheric species except the chosen absorber (either O$_2$ or O$_3$). We calculate the signal by taking the difference between the two spectra: the signal is higher where the absorber has a stronger absorption feature compared to the continuum. We then divide the signal by the noise that PSG simulates based on the chosen instrument to obtain the SNR as a function of wavelength, which is positive only at wavelengths where the absorbing gas absorbs. The total SNR for O$_2$ or O$_3$ is then the square root of the sum of the square of the SNRs \citep{lustig2019detectability}:
\begin{equation}
\text{SNR}_{total} = \sqrt{\sum_{\lambda_i}{(S_{\lambda_i}/N_{\lambda_i})^2}}
\end{equation}
where S$_{\lambda_i}$ and N$_{\lambda_i}$ are the signal and the noise for each wavelength. We calculate the SNRs at 5, 10, 15, and 20 pc, and interpolate between those values for other distances between 5 and 20 pc. For distances below 5 pc or above 20 pc, we scale the SNRs as being inversely proportional to distance \citep{stark2014maximizing}. To get a better idea of whether O$_2$ and O$_3$ would be detectable in real observations for our different assumptions, we first calculate the integration time necessary for the total SNR to be equal to 5.0 for an Earth-like planet at 5, 10, 15, and 20 pc (see Section~\ref{sec:inttimes}). In our Monte Carlo realizations, we consider the observation of an EEC of a given age and at a given distance as a positive life detection if the SNR of either O$_2$ or O$_3$ is above the threshold of 5.0 (see Section~\ref{sec:likelihood}), following \citet{lustig2019detectability}. 

\section{Results}
\label{sec:results}

\subsection{EEC yield required to avoid mission-level false negative scenarios}
\label{sec:theory}
\begin{figure*}
\begin{center}
  \includegraphics[width=1.\textwidth]{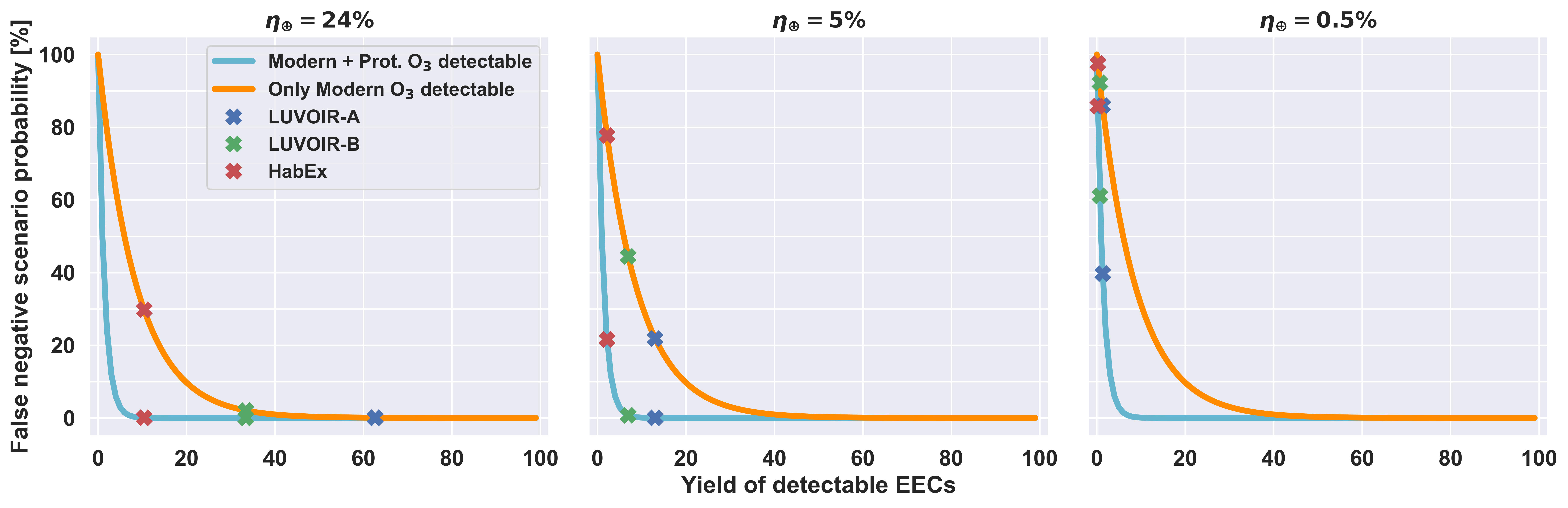}
 \end{center}
  \caption{\textbf{Probability of a mission-level false negative scenario as a function of yield of detectable EECs for three different values of $\eta_{\earth}$. For $\eta_{\earth}$ = 24\%, LUVOIR-A and LUVOIR-B have a $<$1\% and $<$5\% chance of a mission-level false negative scenario, respectively, even if they could only detect modern levels of O$_3$. On the other hand, HabEx needs to be able to detect Proterozoic levels of O$_3$ on every EEC to have $<$1\% chance of a mission-level false negative scenario. For $\eta_{\earth}$ = 5\%, LUVOIR-A and LUVOIR-B must be able to detect Proterozoic levels of O$_3$ to have $<$ 1\% chance of a false negative scenario, while HabEx has $>$20\% chance of one regardless of its ability to detect O$_3$. For $\eta_{\earth}$ = 0.5\%, all missions have probable mission-level false negative scenarios: $\geq$40\%, $\geq$60\%, and $\geq$85\%, for LUVOIR-A, LUVOIR-B, and HabEx, respectively.} Orange line: Only modern levels of O$_3$ are detectable on every EEC; Cyan line: Modern and Proterozoic levels of O$_3$ are detectable on every EEC. Blue cross: LUVOIR-A's mean yield, Green cross: LUVOIR-B's mean yield, Red: HabEx's mean yield.}
  \label{fig:theory}
\end{figure*}
We calculate the probability of a mission-level false negative scenario for an imaginary mission with any yield of EECs. Let's consider two cases: 1) O$_3$ is detectable on every EEC at every distance, but only at modern levels; and 2) O$_3$ is detectable on every EEC at every distance, at both modern and Proterozoic levels. In case 1), O$_3$ is then detectable on 11\% of detectable EECs, as Earth has spent 11\% of its history in the modern era. In case 2), O$_3$ is detectable on 50.7\% of detectable EECs, as Earth has spent 50.7\% of its history in the Proterozoic and modern eras. 

We can then find the probability of a mission-level false negative scenario using the percentage of detectable EECs with detectable O$_3$ for that imaginary mission and its mean EEC yield:
\begin{equation}
P = (1-p)^N
\end{equation}
where P is the probability of a mission-level false negative scenario, p is the percentage of EECs with detectable O$_3$, and N is the mean yield of EECs. We show the two theoretical curves for cases 1) and 2) in Figure~\ref{fig:theory}. We have also superimposed the mean yields of LUVOIR-A, LUVOIR-B, and HabEx on top of the theoretical curves. Using these curves, we can determine the minimum number of EECs that would need to be detectable to have no chance of a mission-level false negative scenario. To have less than a 1\% chance of a mission-level false negative scenario, in case 1), 39.5 EECs would need to be detectable. In case 2), 6.5 EECs would suffice as O$_3$ would be detectable on a greater percentage of them. Similarly, to have less than a 5\% chance of a mission-level false negative scenario, in case 1), 25.7 EECs would need to be detectable, while in case 2), we would need to detect 4.2 EECs. 

For $\eta_{\earth}$ = 24\%, LUVOIR-A's large EEC yield allows it to have less than a 1\% chance of a mission-level false negative scenario even if it could only detect modern levels of O$_3$. Similarly, LUVOIR-B has less than a 5\% chance of a mission-level false negative scenario even if it could only detect modern levels of O$_3$. On the other hand, HabEx needs to be able to reliably detect O$_3$ at Proterozoic levels to avoid a mission-level false negative scenario. That is due to its smaller EEC yield, which requires it to be able to detect lower levels of O$_3$. 

For $\eta_{\earth}$ = 5\%, LUVOIR-A and LUVOIR-B need to be able to detect Proterozoic levels of O$_3$ to have less than a 1\% chance of a mission-level false negative scenario. HabEx on the other hand will have a $>$ 20\% chance of a mission-level false negative scenario even if it could detect O$_3$ at modern and Proterozoic levels on every EEC. 

For $\eta_{\earth}$ = 0.5\%, LUVOIR-A, LUVOIR-B, and HabEx would have a $\sim$40\%, $\sim$ 60\%, and $\sim$ 85\% chance of a mission-level false negative scenario, respectively, even if they could detect O$_3$ at modern and Proterozoic levels on every EEC. Therefore, mission-level false negative scenarios are unavoidable for such a low estimate of $\eta_{\earth}$.

\subsection{Integration Times to detect O$_2$ and O$_3$ with LUVOIR and HabEx}
\label{sec:inttimes}

We present the integration times required for a 5-$\sigma$ detection of O$_2$ and O$_3$ at 5 pc for cloud-free atmospheres calculated using PSG \citep{villanueva2018planetary} in Table~\ref{tab:SNRs} for LUVOIR-A, LUVOIR-B, HabEx/SS and HabEx/no-SS. The integration times calculated at 10, 15, and 20 pc can also be found in the Appendix. To calculate these integration times, we first calculate the SNRs for O$_2$ and O$_3$ detections using the method outlined in Section~\ref{sec:snrcalc} with an exposure time of 1000 hours. We then calculate what the integration time would have to be for these SNRs to be equal to 5.0:
\begin{equation}
\Delta t = 1000 \text{ [hrs]} \times \left(\frac{5}{\text{SNR}} \right)^2
\end{equation}
 We consider six different O$_2$ levels: Modern (1 PAL), and five Proterozoic estimates (10$^{-1}$ to 10$^{-5}$ PAL). 

\begin{center}
\begin{table*}
\centering
  \caption{\textbf{Integration times [hrs] with LUVOIR-A (15 m), LUVOIR-B (8 m), HabEx/SS (4 m), and HabEx/no-SS (4m) to yield a 5-$\sigma$ detection of O$_2$ and O$_3$ for an Earth-like planet without clouds at 5 pc for six different O$_2$ levels. Integration times calculated at 10, 15, and 20 pc can be found in the Appendix.}}
\label{tab:SNRs}
\centering
\begin{tabular}{l l l l l}
    &  15 m LUVOIR-A & 8 m LUVOIR-B & 4 m HabEx/SS & 4 m HabEx/no-SS \\
\hline
O$_2$ = 1 PAL  &  O$_2$: 0.75 hrs &  O$_2$: 3.31 hrs & O$_2$: 12.8 hrs & O$_2$: 44.2 hrs\\ 
   &  O$_3$: 0.19 hr &  O$_3$: 0.56 hr & O$_3$: 0.45 hr & O$_3$: 17.1 hrs \\ 
O$_2$ = 10$^{-1}$ PAL  &  O$_2$: 4.24 hrs &  O$_2$: 17.43 hrs & O$_2$: 65.9 hrs & O$_2$: 211.7 hrs\\ 
   &  O$_3$: 0.33 hr &  O$_3$: 0.89 hrs & O$_3$: 0.41 hr & O$_3$: 62.8 hrs \\ 
O$_2$ = 10$^{-2}$ PAL  & O$_2$: 41.2 hrs &  O$_2$: 167.9 hrs & O$_2$: 564.1 hrs & O$_2$: 2002.1 hrs\\ 
  &  O$_3$: 0.57 hrs &  O$_3$: 1.48 hrs & O$_3$: 0.46 hr & O$_3$: 546.9 hrs \\ 
O$_2$ = 10$^{-3}$ PAL  & O$_2$: 658.0 hrs &  O$_2$: 2679.8 hrs & O$_2$: 6772.4 hrs & O$_2$: 3.2 $\times 10^4$ hrs \\ 
   & O$_3$: 0.99 hrs &  O$_3$: 2.59 hrs & O$_3$: 0.70 hr & O$_3$: 1463.9 hrs \\ 
O$_2$ = 10$^{-4}$ PAL  & O$_2$: 3.1 $\times 10^4$ hrs &  O$_2$: 1.2 $\times 10^5$ hrs & O$_2$: 1.6 $\times 10^5$ hrs & O$_2$: 1.5 $\times 10^6$ hrs \\ 
   & O$_3$: 4.15 hrs &  O$_3$: 11.1 hrs & O$_3$: 2.22 hrs & O$_3$: 9.5 $\times 10^4$ hrs \\ 
O$_2$ = 10$^{-5}$ PAL  & O$_2$: 2.7 $\times 10^6$ hrs &  O$_2$: 1.1 $\times 10^7$ hrs& O$_2$: 1.0 $\times 10^7$ hrs & O$_2$: 1.3 $\times 10^8$ hrs\\ 
   & O$_3$: 125.7 hrs &  O$_3$: 338.2 hrs & O$_3$: 58.94 hrs & O$_3$: 2.0 $\times 10^7$ hrs \\ 
\end{tabular}
\end{table*}
\end{center}

In all cases for LUVOIR and HabEx, O$_3$ is easier to detect than O$_2$ due to its deep and broad feature between 0.2-0.3 $\micron$ (Hartley bands) and between 0.5-0.7 $\micron$ (Chappuis bands). Most importantly, O$_3$ is detectable at 5-$\sigma$ with LUVOIR-A, LUVOIR-B, and HabEx/SS in under 500 hours even at very low estimates for Proterozoic O$_2$ levels (down to 10$^{-5}$ PAL). For HabEx/no-SS, detecting O$_3$ without a starshade at 5 pc is difficult ($>$1000 hours) for Proterozoic O$_2$ levels below 10$^{-2}$ PAL. We note here that the HabEx/SS integration times we calculated for O$_2$ are $\sim$1-4 orders of magnitude larger than reported in the HabEx report (Figure 3.3-7 of \citet{gaudi2020habitable}) depending on the O$_2$ level due to the fact that our background atmosphere includes other absorbers such as O$_3$ and H$_2$O while that of the HabEx report only includes N$_2$. However, our calculations agree closely with simulations made using the \citet{robinson2016characterizing} model when including additional background gases in their model (see Appendix for further details).    

\subsection{Likelihood of detecting O$_2$ and/or O$_3$ with LUVOIR and HabEx}
\label{sec:likelihood}

\subsubsection{$\eta_{\earth}$ = 24\%}
\label{sec:likelihood24}

\begin{figure*}[h!]
\begin{center}
  \includegraphics[width=1.\textwidth]{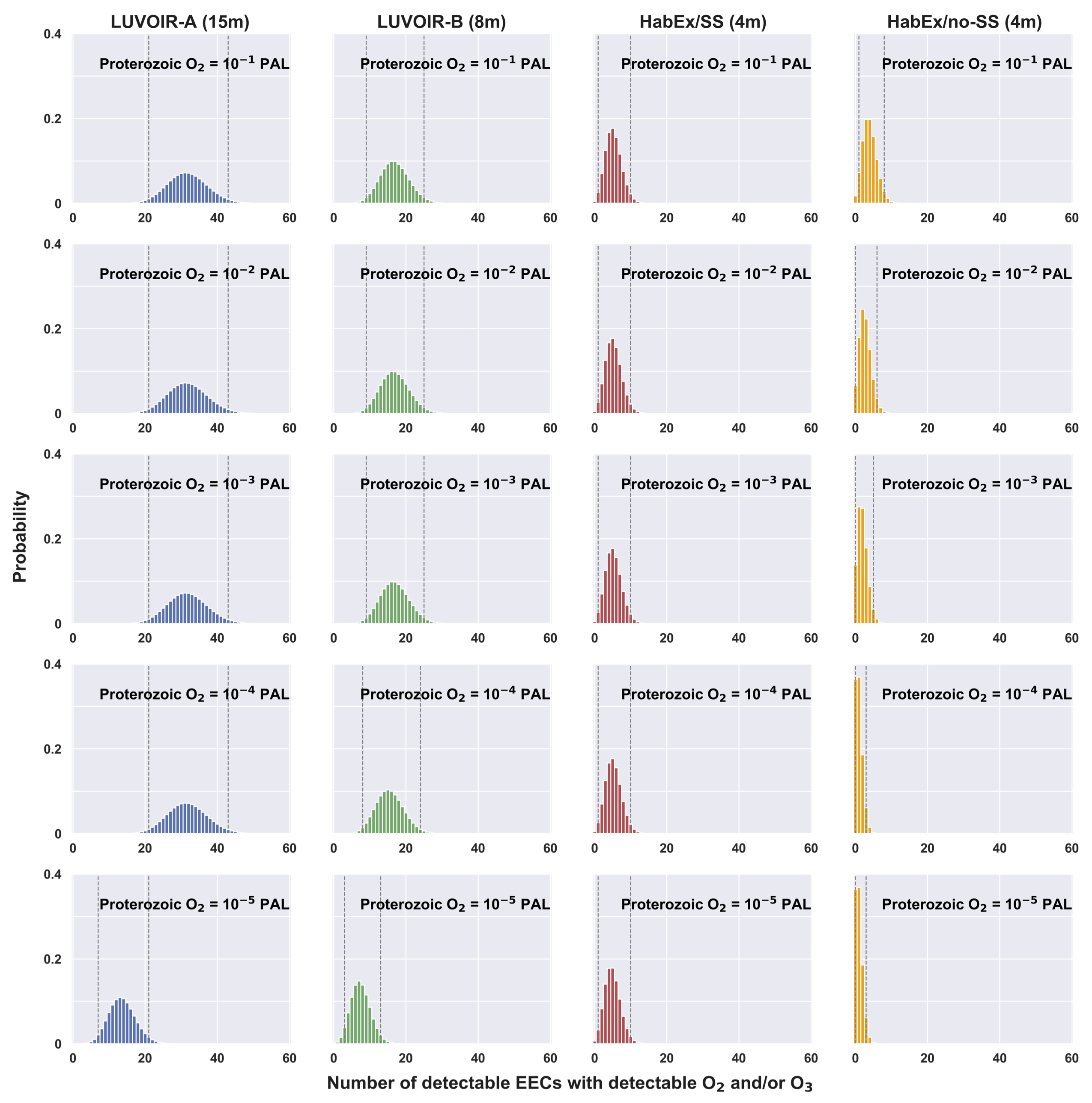}
  \caption{\textbf{Probability distributions of the number of EECs on which an O$_2$ and/or O$_3$ signature could be detected, for $\eta_{\earth}$ = 24\%, assuming f$_E$ = 1. For LUVOIR-A and LUVOIR-B, there are no mission-level false negative scenarios and O$_2$ and/or O$_3$ will be detected on a number of EECs if they are all Earth-like. For HabEx/SS, there is a small mission-level false negative scenario probability (0.5-0.6\%). For HabEx/no-SS, there are different mission-level false negative scenarios depending on the level of O$_2$ we assume for the Proterozoic (probability of 1.8\% for 10$^{-1}$ PAL, 6.6\% for 10$^{-2}$ PAL, 13.8\% for 10$^{-3}$ PAL, and 36.4\% for $\leq$10$^{-4}$ PAL of Proterozoic O$_2$).} Panels: 15 m segmented on-axis LUVOIR-A (left-most), an 8 m segmented off-axis LUVOIR-B (center left), 4 m monolith off-axis HabEx/SS (starshade, center right), and 4 m monolith off-axis HabEx/no-SS (no starshade, right-most). We assume that all EECs are Earth-like (f$_E$ = 1). If the distribution reaches zero, there is a non-zero mission-level false negative probability: O$_2$ and/or O$_3$ are not detected on any EECs despite all of them being Earth-like. We draw an age for EECs between Hadean, Archean, pre-GOE Proterozoic, post-GOE Proterozoic, and Modern eras, and assume Hadean, Archean, and pre-GOE Proterozoic O$_2$ and O$_3$ levels are undetectable. Post-GOE Proterozoic O$_2$ concentrations range between 10$^{-5}$ and 0.1 PAL. The vertical grey dotted lines represent the 95\% confidence interval.}
  \label{fig:summary_nc}
\end{center}
\end{figure*}

We present the probability distributions of total O$_2$ and/or O$_3$ detections for LUVOIR-A (left-most), LUVOIR-B (center-left), HabEx/SS (center right), and HabEx/no-SS (right-most) in Figure~\ref{fig:summary_nc}, assuming $\eta_{\earth}$ = 24\%. As described in Section~\ref{sec:methods}, we calculate the SNR of O$_2$ and O$_3$ for a given EEC at a certain distance and with a certain level of O$_2$ based on its age drawn from a uniform distribution. If either of these SNRs is above the threshold of 5.0 (for an exposure time of 1000 hours), we consider that EEC a positive O$_2$/O$_3$ detection. We note here that since O$_3$ is easier to detect than O$_2$ in all cases we considered (see Table~\ref{tab:SNRs}), O$_3$ detectability is the limiting factor in determining whether O$_2$ and/or O$_3$ is detectable on a given EEC. We assume that f$_E$ = 1 (all EECs are Earth-like), and determine whether there are any mission-level false negative scenario even under this optimistic assumption. We vary the Proterozoic O$_2$ from 10$^{-5}$ to 0.1 PAL.

For LUVOIR-A, with Proterozoic levels of O$_2$ equal to or larger than 10$^{-4}$ PAL, O$_3$ is detectable for EECs at all target distances for both modern and Proterozoic levels. Because of this, the first four rows of Figure~\ref{fig:summary_nc} show the same distribution. The reason the distribution is offset from 62 (peak of Figure~\ref{fig:Np}'s left panel) is that O$_2$ and O$_3$ are undetectable for EECs in a Hadean or Archean era and so all EECs drawn to be of Hadean or Archean age will correspond to a null detection of O$_2$ and O$_3$. For these four levels of Proterozoic O$_2$, we find that LUVOIR-A has a 95\% chance of detecting 21-43 Earth-like EECs (for f$_E = 1$). At Proterozoic O$_2$ levels of 10$^{-5}$ PAL, O$_3$ is only detectable on EECs at close distances ($<$11.5 pc), and therefore the distribution shifts toward zero due to the EECs that are further away and drawn to be of Proterozoic age. We find that LUVOIR-A has a 95\% chance of detecting 7-21 Earth-like EECs for Proterozoic levels of 10$^{-5}$ PAL (for f$_E = 1$). In the first four cases (Proterozoic O$_2$ between 0.1-10$^{-4}$ PAL), the distribution of the number of EECs that have detectable O$_3$ is well above zero. For the case with Proterozoic levels of 10$^{-5}$ PAL, the curve's end member is above zero with 5 as the smallest number of detected Earth-like EECs. There is therefore no mission-level false negative scenario for LUVOIR as a mission if f$_E = 1$: O$_2$ and/or O$_3$ will be detected on a number of EECs if they are generally Earth-like.

For LUVOIR-B, with Proterozoic levels of O$_2$ equal to or larger than 10$^{-3}$ PAL, O$_3$ is detectable for EECs at all target distances for both modern and Proterozoic levels. At lower Proterozoic O$_2$ levels, O$_3$ is only detectable at $<$19.1 pc for 10$^{-4}$ PAL and at $<$8.1 pc for 10$^{-5}$ PAL. We find that LUVOIR-B has a 95\% chance (for f$_E = 1$) of detecting 9-25 Earth-like EECs for Proterozoic O$_2$ levels of 0.1-10$^{-3}$ PAL, 8-24 for 10$^{-4}$ PAL, and 3-13 for 10$^{-5}$ PAL. Similarly to LUVOIR-A, for Proterozoic O$_2$ between 0.1-10$^{-4}$ PAL, the distribution is well above zero, while for 10$^{-5}$ PAL, the smallest number of detected Earth-like EECs is 1. Therefore, similarly to LUVOIR-A there are no mission-level false negative scenarios for LUVOIR-B as a mission if f$_E = 1$.
\begin{table*}
  \caption{\textbf{Probability of a mission-level false negative scenario for different assumptions of Proterozoic O$_2$ concentrations, assuming f$_E$ = 1, for $\eta_{\earth}$ = 24\%.} We draw a distance from the mission target list and an age for each EEC observed. A mission-level false negative scenario is defined as not detecting O$_2$/O$_3$ on any of the observed EECs even though we assume they are all Earth-like. The lowest possible mission-level false negative probability we can estimate is $<$ 10$^{-5}$\% given that we perform 10$^7$ Monte Carlo simulations.}
\label{tab:falseneg24}
\centering
\begin{tabular}{l|llll}
     &  15 m LUVOIR-A & 8 m LUVOIR-B & 4 m HabEx/SS & 4 m HabEx/no-SS \\
\hline
O$_2$ = 10$^{-1}$ PAL  & $<$10$^{-5}$\% & $<$10$^{-5}$\% & 0.48\% & 1.8\% \\
O$_2$ = 10$^{-2}$ PAL  &  $<$10$^{-5}$\% & $<$10$^{-5}$\% & 0.48\% & 6.6\%  \\
O$_2$ = 10$^{-3}$ PAL  &  $<$10$^{-5}$\% & $<$10$^{-5}$\% & 0.48\% & 13.8\% \\
O$_2$ = 10$^{-4}$ PAL  &  $<$10$^{-5}$\% & $<$10$^{-5}$\% & 0.48\% & 36.4\% \\
O$_2$ = 10$^{-5}$ PAL  &  6.0$\times$10$^{-5}$\% & 0.054\% & 0.63\% & 36.4\% \\
\end{tabular}
\end{table*}

\begin{table*}
  \caption{\textbf{Same as Table~\ref{tab:falseneg24} but for $\eta_{\earth}$ = 5\%.}}
\label{tab:falseneg5}
\centering
\begin{tabular}{l|llll}
     &  15 m LUVOIR-A & 8 m LUVOIR-B & 4 m HabEx/SS & 4 m HabEx/no-SS \\
\hline
O$_2$ = 10$^{-1}$ PAL  &  0.13\% & 2.9\% & 33.3\% & 43.6\% \\
O$_2$ = 10$^{-2}$ PAL  &  0.13\% & 2.9\% & 33.3\% & 56.9\% \\
O$_2$ = 10$^{-3}$ PAL  &  0.13\% & 2.9\% & 33.3\% & 66.3\% \\
O$_2$ = 10$^{-4}$ PAL  & 0.13\% & 3.9\% & 33.3\% & 81.1\% \\
O$_2$ = 10$^{-5}$ PAL  &  5.8\% & 21.0\% & 35.3\% & 81.1\% \\
\end{tabular}
\end{table*}

\begin{table*}
  \caption{\textbf{Same as Table~\ref{tab:falseneg24} but for $\eta_{\earth}$ = 0.5\%.}}
\label{tab:falseneg05}
\centering
\begin{tabular}{l|llll}
     &  15 m LUVOIR-A & 8 m LUVOIR-B & 4 m HabEx/SS & 4 m HabEx/no-SS \\
\hline
O$_2$ = 10$^{-1}$ PAL  &  51.6\% & 70.3\% & 89.6\% & 91.0\% \\
O$_2$ = 10$^{-2}$ PAL  &  51.6\% & 70.3\% & 89.6\% & 94.5\%  \\
O$_2$ = 10$^{-3}$ PAL  &  51.6\% & 70.3\% &89.6\% & 96.0\%  \\
O$_2$ = 10$^{-4}$ PAL  &  51.6\% & 72.4\% & 89.6\% & 97.9\%  \\
O$_2$ = 10$^{-5}$ PAL  &  75.3\% & 85.6\% & 90.1\% & 97.9\%  \\
\end{tabular}
\end{table*}
For HabEx/SS, with Proterozoic levels of O$_2$ equal to or larger than 10$^{-4}$ PAL, O$_3$ is detectable for EECs at all target distances for both modern and Proterozoic levels. At 10$^{-5}$ PAL, O$_3$ is only detectable on EECs at $<$13.9 pc. We find that HabEx/SS has a 95\% chance of detecting 1-10 Earth-like EECs for all levels of Proterozoic O$_2$ considered (for f$_E = 1$). Because of the small number of EECs that HabEx can detect, the smallest possible number of Earth-like EECs for HabEx is 0. There is therefore a probability of a mission-level false negative scenario with HabEx even if f$_E = 1$, however it is only 0.5-0.6\%. 

\begin{figure*}
\centering
  \includegraphics[width=1.\textwidth]{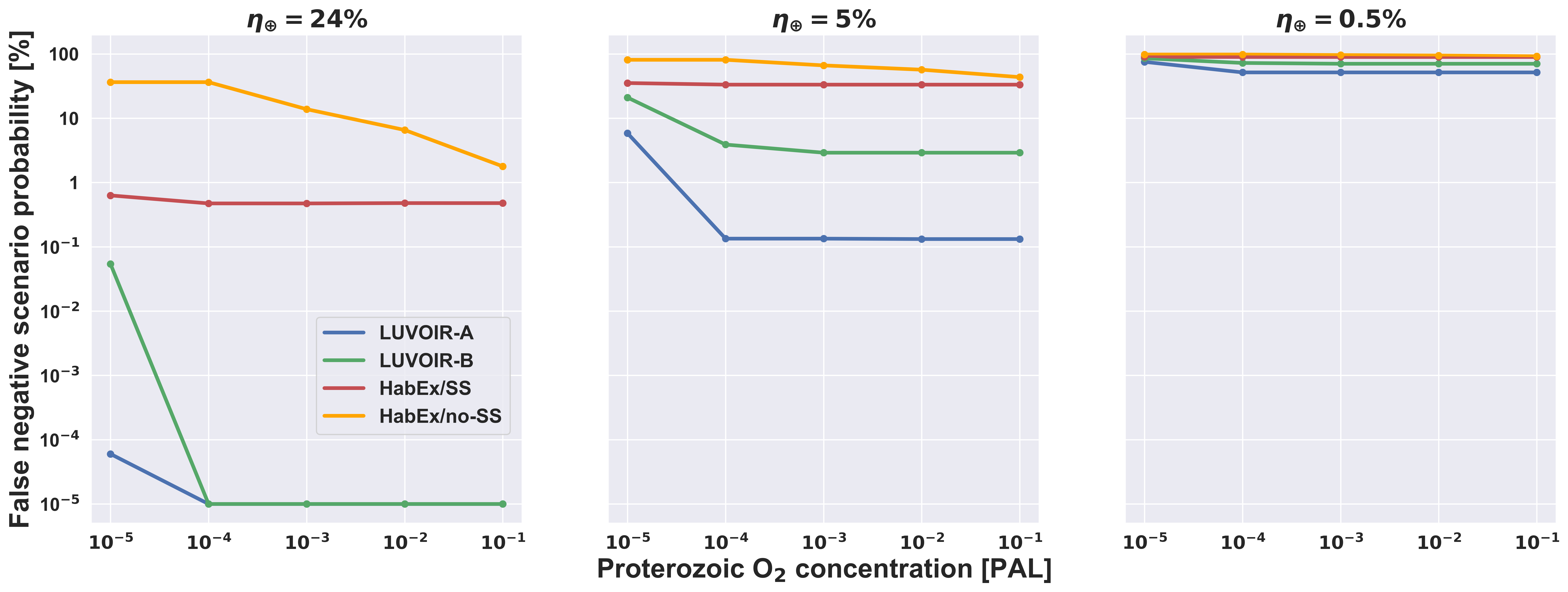}
  \caption{\textbf{Probability of a mission-level false negative scenario for different assumptions of Proterozoic O$_2$ concentrations, assuming f$_E$ = 1, for $\eta_{\earth}$ = 24\%, 5\%, and 0.5\%. LUVOIR-A, given its large expected EEC yield, is most robust to lowering estimates of $\eta_{\earth}$, allowing low probabilities of a false negative scenario for an $\eta_{\earth}$ estimate down to 5\%.} A mission-level false negative scenario is defined as not detecting O$_2$ or O$_3$ on any of the observed EECs even though we assume they are all Earth-like. The lowest possible mission-level false negative probability we can estimate is $<$10$^{-5}$\% given that we perform 10$^7$ Monte Carlo simulations.}
  \label{fig:falsenegs}
\end{figure*}

For HabEx/no-SS, O$_3$ is only detectable at certain distances for high enough levels of Proterozoic O$_2$: $<$13.1 pc for 1 PAL, $<$9.8 pc for 10$^{-1}$ PAL,$<$6.6 pc for 10$^{-2}$ PAL, $<$4.1 pc for 10$^{-3}$ PAL. At Proterozoic O$_2$ levels of 10$^{-4}$ and 10$^{-5}$ PAL, O$_3$ is not detectable at any distance. We find that HabEx/no-SS has a 95\% chance (for f$_E = 1$) of detecting 1-8 Earth-like EECs for Proterozoic levels of 0.1 PAL, 0-6 for 10$^{-2}$ PAL, 0-5 for 10$^{-3}$ PAL, and 0-3 for 10$^{-4}$ and 10$^{-5}$ PAL. For all cases of Proterozoic O$_2$, HabEx/no-SS has a mission-level false negative scenario where we do not detect O$_2$ or O$_3$ on any of the EECs we observe even though f$_E = 1$. That mission-level false negative probability depends on the assumed Proterozoic O$_2$ level, and in the cases we considered it is: 1.8\% for 10$^{-1}$, 6.6\% for 10$^{-2}$, 13.8\% for 10$^{-3}$, and 36.4\% for 10$^{-4}$ and 10$^{-5}$ PAL. We summarize these mission-level false negative probabilities in Table~\ref{tab:falseneg24} and Figure~\ref{fig:falsenegs}.

\subsubsection{$\eta_{\earth}$ = 5\%}
\label{sec:likelihood5}

\begin{figure*}[h!]
\begin{center}
  \includegraphics[width=1.\textwidth]{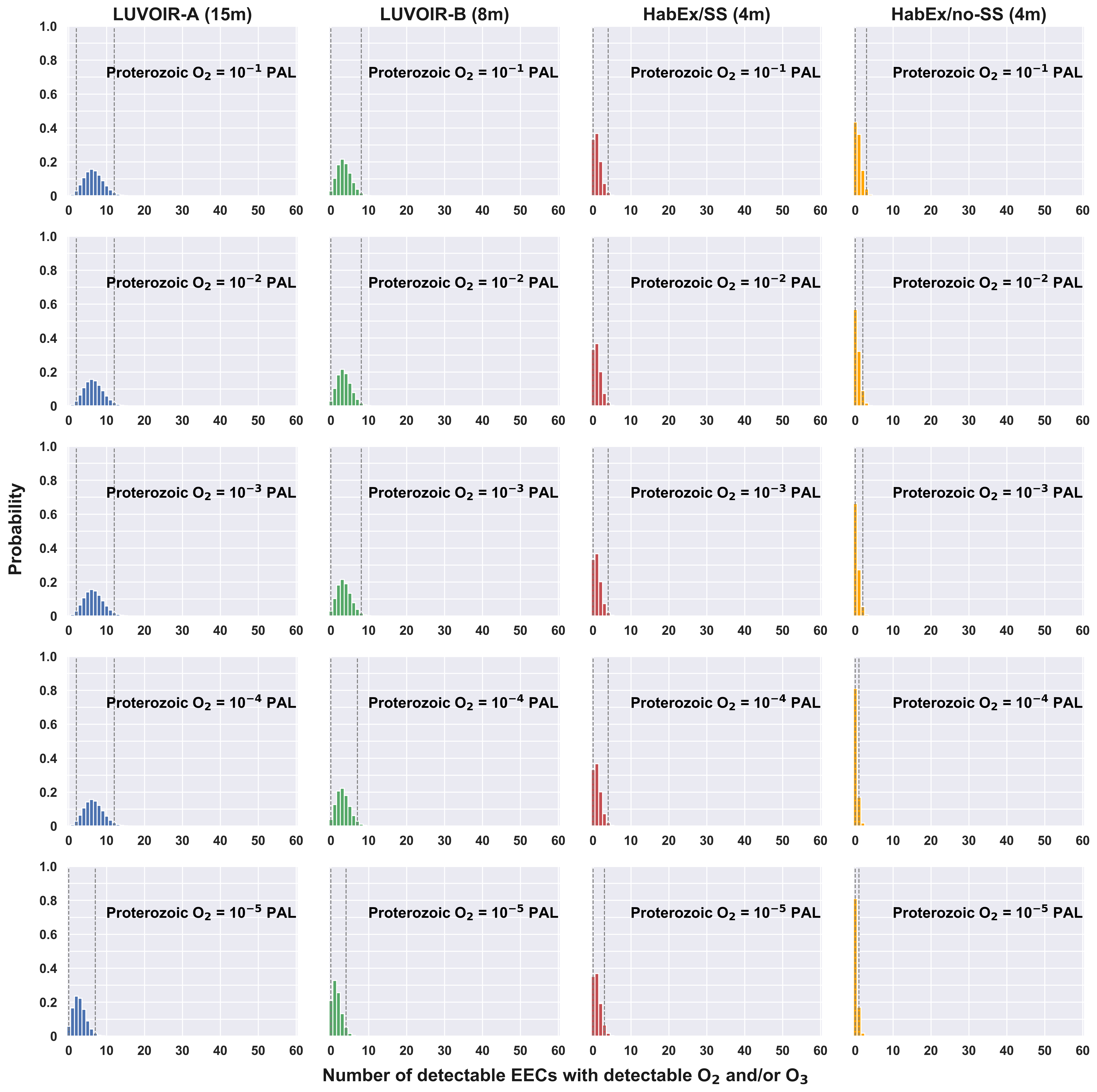}
  \caption{\textbf{Same as Figure~\ref{fig:summary_nc} but for $\eta_{\earth}$ = 5\%, assuming f$_E$ = 1. For LUVOIR-A, there are no mission-level false negative scenarios for Proterozoic O$_2$ $\geq$ 10$^{-4}$ PAL. For Proterozoic O$_2$ of 10$^{-5}$ PAL there is a 5\% chance of a mission-level false negative scenario. For LUVOIR-B, there is a 1.4\% chance of a mission-level false negative scenario for Proterozoic O$_2$ $\geq$ 10$^{-3}$ PAL, a 2.3\% chance for 10$^{-4}$ PAL, and a 20.2\% chance for 10$^{-5}$ PAL. For HabEx/SS, there is a 34.2\%  chance of a mission-level false negative scenario for Proterozoic O$_2$ $\geq$ 10$^{-4}$ PAL, and a 36.7\% chance for 10$^{-5}$ PAL. For HabEx/no-SS, the mission-level false negative probabilities are as follows: 46.9\% for 10$^{-1}$, 61.6\% for 10$^{-2}$, 71\% for 10$^{-3}$, and 84.5\% for $\leq$10$^{-4}$ PAL of Proterozoic O$_2$.} Note that the scale of the y-axis is different than in Figure~\ref{fig:summary_nc}.}
  \label{fig:summary_nc5}
\end{center}
\end{figure*}

For the lower $\eta_{\earth}$ estimate of 5\%, Figure~\ref{fig:summary_nc5} shows that all distributions are shifted toward zero compared to Figure~\ref{fig:summary_nc}. The most important difference compared to Figure~\ref{fig:summary_nc} is the existence and greater likelihood of mission-level false negative scenarios for f$_E = 1$. LUVOIR-A still has negligible ($\sim$ 0.13\%) mission-level false negative scenarios for Proterozoic O$_2$ levels $\geq$10$^{-4}$ PAL, but has a mission-level false negative probability of 5.8\% for Proterozoic O$_2$ of 10$^{-5}$ PAL, for f$_E = 1$. LUVOIR-B has mission-level false negative scenario probabilities for every Proterozoic O$_2$ case (2.9\% for $\geq$ 10$^{-3}$, 3.9\% for 10$^{-4}$, and 21.0\% for 10$^{-5}$ PAL), for f$_E = 1$. HabEx/SS and HabEx/no-SS have significantly greater mission-level false negative scenario probabilities. HabEx/SS has a 33.3\% chance of a mission-level false negative probability for Proterozoic O$_2$ $\geq$10$^{-4}$ and 35.3\% for 10$^{-5}$ PAL, for f$_E = 1$. HabEx/no-SS has a 43.6\% chance of a mission-level false negative for Proterozoic O$_2$ of 10$^{-1}$, 56.9\% for 10$^{-2}$, 66.3\% for 10$^{-3}$, and 81.1\% for $\leq$10$^{-4}$ PAL, for f$_E = 1$. We summarize these mission-level false negative probabilities in Table~\ref{tab:falseneg5} and Figure~\ref{fig:falsenegs}.

\subsubsection{$\eta_{\earth}$ = 0.5\%}
\label{sec:likelihood05}

The lowest estimate of $\eta_{\earth}$ predicts that EECs are rare enough that it will be difficult to detect them in the first place (see Figure~\ref{fig:Np}'s right panel), as there is a 27\%, 50\%, and 81\% chance of not detecting any of them for LUVOIR-A, LUVOIR-B, and HabEx, respectively. Because of this, for f$_E = 1$, there is a 51.6\% chance of a mission-level false negative scenario with LUVOIR-A for Proterozoic O$_2$ $\geq$ 10$^{-4}$ PAL and 75.3\% for Proterozoic O$_2$ of 10$^{-5}$ PAL. LUVOIR-B has a 70.3\% chance of a false negative for Proterozoic O$_2$ $\geq$ 10$^{-3}$ PAL, 72.4\% for 10$^{-4}$ PAL, and 85.6\% for 10$^{-5}$ PAL, for f$_E = 1$. HabEx/SS has a 89.6\% chance of a false negative for Proterozoic O$_2$ $\geq$ 10$^{-4}$ PAL and 90.1\% chance for Proterozoic O$_2$ of 10$^{-5}$ PAL, for f$_E = 1$. HabEx/no-SS has a 91.0\%, 94.5\%, 96.0\%, and 97.9\% chance of a false negative for Proterozoic O$_2$ of 10$^{-1}$ PAL, 10$^{-2}$ PAL, 10$^{-3}$ PAL, and $\leq$ 10$^{-4}$ PAL, for f$_E = 1$. To summarize, if $\eta_{\earth}$ = 0.5\%, all missions that we considered will detect too few EECs to rule out false negative scenarios even if f$_E = 1$. We do not include a reproduction of Figures~\ref{fig:summary_nc} and~\ref{fig:summary_nc5} for $\eta_{\earth}$ = 0.5\% as most panels simply show high peaks at zero, but we summarize the mission-level false negative probabilities in Table~\ref{tab:falseneg05} and Figure~\ref{fig:falsenegs}.

\subsection{Using a null detection to constrain the fraction f$_E$ of Earth-like EECs}
\label{sec:fraction}

In the highly likely scenario that only a fraction $f_E$ of EECs are actually Earth-like, the number of planets on which we detect O$_2$/O$_3$ will be decreased by a factor $f_E$. For example, if only 10\% of EECs are Earth-like ($f_E = 0.1$), assuming $\eta_{\earth}$ = 24\%, the peak of Figure~\ref{fig:summary_nc}'s first row panel for LUVOIR-A would shift from 31 to 3.1. For HabEx/SS the peak of all panels would shift toward zero to create distributions with peaks at zero, making it unlikely that we will be able to detect O$_2$ or O$_3$ with HabEx/SS if $f_E$ is low. 

An instrument that lacks mission-level false negatives for $f_E = 1$, such as (for $\eta_{\earth}$ = 24\%) LUVOIR-A, LUVOIR-B, or HabEx/SS, will allow our observations to put a constraint on f$_E$ even if we cannot detect O$_2$ or O$_3$. On the other hand, an instrument such as HabEx/no-SS will not allow us to make any inference about f$_E$ in the event of a null detection, no matter the value of $\eta_{\earth}$, as it could be caused by a mission-level false negative scenario even when $f_E = 1$. For missions lacking mission-level false negative scenarios when $f_E = 1$, how well we can constrain f$_E$ depends on the mission and on the value of $\eta_{\earth}$, and is a function of the number of detectable EECs and of the ability to detect various levels of O$_3$. 

\begin{figure*}
\centering
  \includegraphics[width=1.\textwidth]{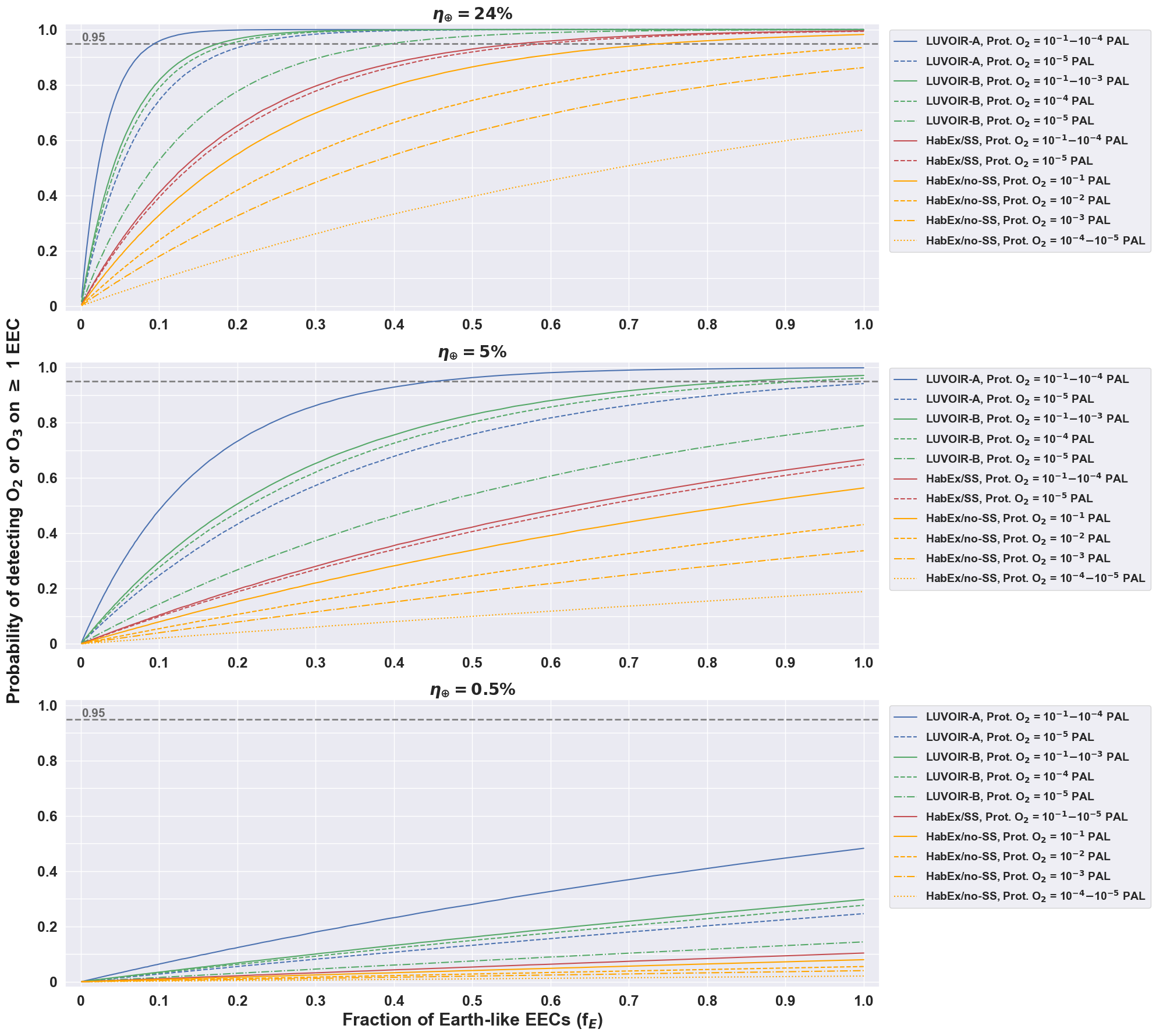}
  \caption{\textbf{Probability of detecting O$_2$ or O$_3$ on at least 1 EEC as a function of the fraction of Earth-like EECs (f$_E$), for $\eta_{\earth}$ = 24\%, 5\%, and 0.5\%. The intersection of each curve with the 95\% vertical grey line denotes the upper limit that will be placed on f$_E$ at 95\% confidence in the event of a null detection of O$_2$ and O$_3$ on all observed EECs.} If the curve does not intersect the 95\% grey line, a constraint cannot be placed on f$_E$ in the event of a null detection due to the existence of mission-level false negative scenarios. For a given Proterozoic O$_2$ level, the mission with the highest expected number of detectable EECs allows for the best constraint to be placed on f$_E$. }
  \label{fig:fEfig}
\end{figure*}

For different values of f$_E$, we perform Monte Carlo simulations by resampling from the distributions in Figure~\ref{fig:summary_nc} and weighing each draw by f$_E$, resulting in the number of EECs with detectable O$_2$ or O$_3$ for that value of f$_E$. We show the probability that at least 1 Earth-like EEC will have detectable O$_2$ or O$_3$ as a function of the fraction fo Earth-like EECs, f$_E$, in Figure~\ref{fig:fEfig}.

\subsubsection{$\eta_{\earth}$ = 24\%}
\label{sec:constrain24}

If we do not detect O$_2$ or O$_3$ on any EECs with LUVOIR-A, this null detection would mean that f$_E$ $\leq$0.094 for Proterozoic O$_2$ levels above 10$^{-4}$ PAL, and $\leq$0.22 for Proterozoic O$_2$ levels of 10$^{-5}$ PAL. Therefore, a null detection with LUVOIR-A will constrain f$_E$ with 95\% confidence to $\leq 0.094 - 0.22$. Similarly, a null result with LUVOIR-B would mean that f$_E$ is $\leq$0.18 for Proterozoic O$_2$ levels above 10$^{-3}$ PAL, $\leq$0.19 for Proterozoic O$_2$ levels of 10$^{-4}$ PAL, and $\leq$0.40 for Proterozoic O$_2$ levels of 10$^{-5}$ PAL. Therefore, a null detection with LUVOIR-B will constrain f$_E$ $\leq 0.18 - 0.40$ with 95\% confidence. HabEx/SS will also allow us to constrain f$_E$ $\leq 0.56$ for Proterozoic O$_2$ levels above 10$^{-4}$ PAL, and $\leq$0.59 for Proterozoic O$_2$ levels of 10$^{-5}$ PAL in the case of a null detection, all with 95\% confidence. On the other hand, a null detection with HabEx/no-SS will not allow us to constrain f$_E$ as there are mission-level false negative scenarios where we do not detect O$_2$ or O$_3$ even if all EECs are Earth-like. These potential constraints on f$_E$ are summarized in Figure~\ref{fig:constraint}. 

This also implies that we have a 95\% chance of detecting O$_2$ or O$_3$ on at least 1 EEC with LUVOIR-A for $f_E > 0.094$, with LUVOIR-B for $f_E > 0.18$, and with HabEx/SS for $f_E > 0.56$ (all for Proterozoic O$_2$ $\geq$ 10$^{-4}$ PAL, which is the likely lower limit of the Proterozoic O$_2$ range). However, a null detection is possible with HabEx/no-SS for any value of $f_E$. 

\subsubsection{$\eta_{\earth}$ = 5\%}
\label{sec:constrain5}

\begin{figure*}
\centering
  \includegraphics[width=1.\textwidth]{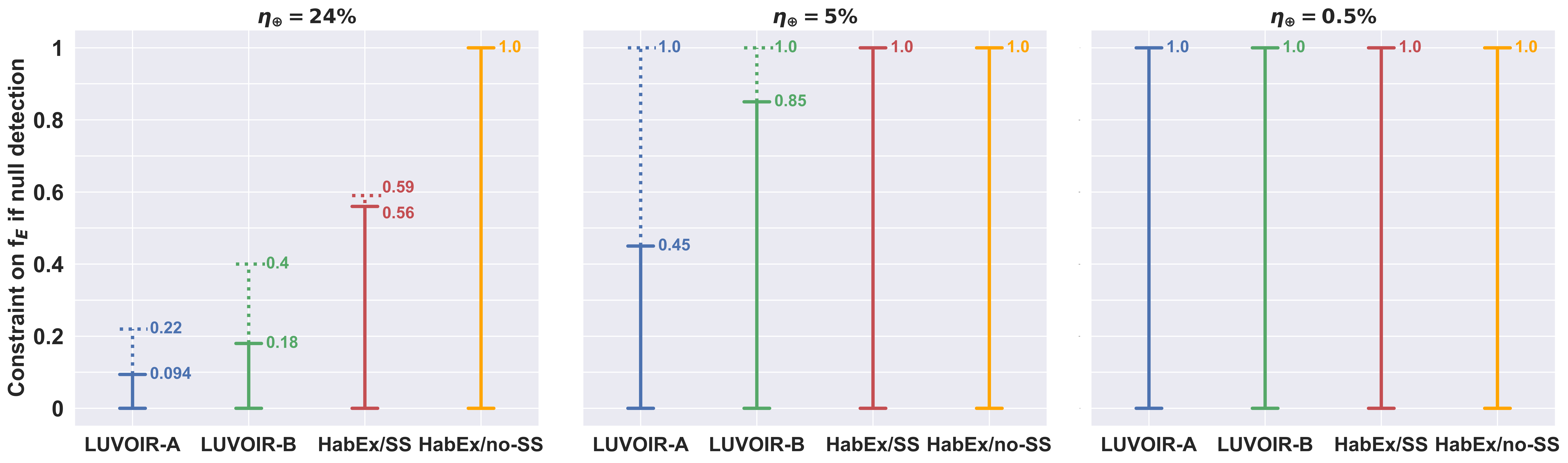}
  \caption{\textbf{Constraint that we will be able to place on f$_E$, with 95\% confidence, in the case of a null detection where we do not detect O$_2$ or O$_3$ on any EECs for $\eta_{\earth}$ = 24\%, 5\%, and 0.5\%. These constraints also imply that for f$_E$ greater than the upper limits, we have a 95\% chance to detect O$_3$ on at least 1 EEC.} Solid lines: constraints for Proterozoic O$_2$ level of 10$^{-1}$$-$10$^{-4}$ PAL; Dashed lines: constraints for Proterozoic O$_2$ level of 10$^{-5}$ PAL. An upper limit of 1.0 for the constraint on f$_E$ means that there is a $\geq$ 5\% chance of a mission-level false negative scenario, where we do not detect O$_2$ or O$_3$ on any EECs even if they are all Earth-like, and so f$_E$ cannot be constrained. Note that LUVOIR-B's constraint on f$_E$ will be $\leq 19$\% for $\eta_{\earth}$ = 24\% and Proterozoic O$_2$ level of 10$^{-4}$ PAL, and $\leq 92$\% for $\eta_{\earth}$ = 5\% and Proterozoic O$_2$ level of 10$^{-4}$ PAL.}
  \label{fig:constraint}
\end{figure*}

A lower $\eta_{\earth}$ estimate of 5\% would also impact the extent to which we could constrain f$_E$ in the case of a null detection. If we do not detect O$_2$ or O$_3$ on any EECs with LUVOIR-A, we could constrain f$_E$ with 95\% confidence to $\leq$ 0.45 for Proterozoic O$_2$ $\geq$10$^{-4}$ PAL. However, for Proterozoic O$_2$ of 10$^{-5}$ PAL, we will not be able to constrain f$_E$ as there is a 6\% chance of a mission-level false negative scenario. Similarly, a null detection with LUVOIR-B will allow us to constrain f$_E$ with 95\% confidence to $\leq$ 0.85 for Proterozoic O$_2$ $\geq$10$^{-3}$ PAL, and to  $\leq$ 0.92 for Proterozoic O$_2$ of 10$^{-4}$ PAL, but we will not be able to constrain it at lower Proterozoic levels as there is a 21\% chance of a mission-level false negative for Proterozoic O$_2$ of 10$^{-5}$ PAL. HabEx/SS and HabEx/no-SS have a $33-35$\% and $44-81$\% chance of a mission-level false negative, respectively, and therefore will not allow us to constrain f$_E$. These potential constraints on f$_E$ are summarized in Figure~\ref{fig:constraint}. 

This also implies that we have a 95\% chance of detecting O$_2$ or O$_3$ on at least 1 EEC with LUVOIR-A for $f_E > 0.45$, with LUVOIR-B for $f_E > 0.85$. However, a null detection is possible with HabEx/SS and HabEx/no-SS for any value of $f_E$.

\subsubsection{$\eta_{\earth}$ = 0.5\%}
\label{sec:constrain05}

Using the lowest estimate of $\eta_{\earth}$ = 0.5\% introduces the possibility of detecting no EECs in the first place, hence a null detection of O$_2$ and O$_3$ on all EECs is unlikely to help constrain f$_E$. In this scenario, none of the missions would inform us on the likely value of f$_E$ in the case of a null detection, and it is likely we would not detect O$_2$ or O$_3$ on any EECs even if they are all Earth-like.

\section{Discussion}
\label{sec:discussion}

If EECs are all Earth-like in that they develop Earth-like oxygenic photosynthesis that oxygenates their atmosphere, we should detect O$_2$ or O$_3$ with LUVOIR-A, LUVOIR-B, and HabEx/SS (with starshade) as long as $\eta_{\earth}$ is sufficiently large. For $\eta_{\earth}$ = 24\%, LUVOIR-A and LUVOIR-B have a negligible chance of a mission-level false negative scenario, where EECs are all Earth-like but we do not detect O$_2$ or O$_3$ on any of them. HabEx/SS similarly has only a very small chance of such a false negative. For $\eta_{\earth}$ = 5\%, for Proterozoic O$_2$ $\geq$ 10$^{-4}$ PAL, LUVOIR-A similarly has a negligible chance of a mission-level false negative scenario, and LUVOIR-B's chance of a mission-level false negative scenario is $<$ 4\%, while HabEx/SS has a 33\% chance of one. For $\eta_{\earth}$ = 0.5\%, it is reasonably likely that no EEC will be detected in the first place: there is a 27\%, 50\%, and 81\% chance that LUVOIR-A, LUVOIR-B, and HabEx do not detect any. As a result, all missions have a significant probability of a mission-level false negative scenario even if all EECs are Earth-like (f$_E$ = 1). For any value of $\eta_{\earth}$, HabEx/no-SS has a significant chance of a mission-level false negative scenario, showing therefore that a starshade is crucial to HabEx's design. 

If we detect O$_2$ or O$_3$ on a number of EECs that is lower than the number we would expect if f$_E$ = 1, we will be able to constrain the fraction of EECs that are actually Earth-like (f$_E$) with all instruments. In the event that we do not detect O$_2$ or O$_3$ on any EECs we will also be able to constrain f$_E$ to different extents based on the mission and based on the value of $\eta_{\earth}$. For $\eta_{\earth}$ = 24\%, LUVOIR-A, LUVOIR-B, and HabEx/SS will allow us to constrain f$_E$ $\leq$ 0.094, f$_E$ $\leq$ 0.18, and f$_E$ $\leq$ 0.56, respectively, with 95\% confidence, for Proterozoic O$_2$ levels $\geq$ 10$^{-4}$ PAL. On the other hand, HabEx/no-SS will not allow us to constrain f$_E$ due to the existence of mission-level false negative scenarios. For $\eta_{\earth}$ = 5\%, LUVOIR-A and LUVOIR-B will allow us to constrain f$_E$ $\leq$ 0.45 and f$_E$ $\leq$ 0.85, respectively, while HabEx/SS and HabEx/no-SS will not allow us to constrain f$_E$. For $\eta_{\earth}$ = 0.5\%, none of the missions will allow us to constrain f$_E$ due to the high probability of mission-level false negative scenarios. This illustrates the fact that a mission with a higher expected yield of EECs is more robust to the uncertainty in the value of $\eta_{\earth}$, and that further constraining $\eta_{\earth}$ should be a priority to allow us to make predictions of the outcome of future observations.

As HabEx and LUVOIR are still mission concepts, their specifications are not yet finalized. The number of EECs that we can detect with either instrument depends on the set of specificities we choose from \citet{stark2019exoearth}. Particularly, our analysis was performed using a high throughput scenario for both instruments, which is likely optimistic. We considered a pessimistic scenario where we use a low throughput, and recreated Figure~\ref{fig:Np}. For $\eta_{\earth}$ = 24\%, we found that the number of EECs that we could detect with LUVOIR-A decreases from 62 to 50, with LUVOIR-B from 33 to 27, and with HabEx from 10 to 8. In this low throughput scenario, the peaks of Figure~\ref{fig:summary_nc} are shifted to the left as well. For example, for $\eta_{\earth}$ = 24\% and Proterozoic O$_2$ levels of 10$^{-5}$ PAL, this increases the mission-level false negative scenario probability from 0.6\% (in the high throughput scenario) to 1.8\% for HabEx/SS, and from 36\% to 45\% for HabEx/no-SS. On the other hand, assuming a low throughput does not introduce any mission-level false negative scenario for LUVOIR-A and LUVOIR-B for this estimate of $\eta_{\earth}$.

Exoplanet atmospheres are expected to be diverse in their composition, even if they are inhabited by Earth-like life. In this work, we considered an Earth-like background atmosphere in our SNR calculations that includes gases such as O$_2$, O$_3$, CO$_2$, H$_2$O, N$_2$, and CH$_4$. However, EEC atmospheres may contain different gases that could affect the SNRs of O$_2$ and O$_3$ features. Observations would require a full atmospheric retrieval study to confirm O$_2$ and O$_3$ detections \citep{feng2018characterizing}. 

The HabEx/SS integration times we calculated for O$_2$ are greater than those reported in the HabEx report \citep{gaudi2020habitable}. We find that the integration times necessary to detect O$_2$ and O$_3$ with HabEx/SS are 1-4 orders of magnitude larger than in their study (see their Figure 3.3-7). The reason for this is that we included various gases such as O$_3$, CO$_2$, and H$_2$O in our background atmosphere, while the HabEx report calculations only considered N$_2$ as a background gas. Because of this, in their calculations the difference between the continuum (N$_2$ only) and the absorption spectrum (N$_2$ and O$_2$) is large at $\sim$0.2 $\micron$ as N$_2$ does not absorb there. In Earth-like atmospheres, species such as O$_3$ and H$_2$O absorb at these short wavelengths. This minimizes the difference between the continuum spectra (without O$_2$) and the absorber spectra (with O$_2$) at 0.2 $\micron$. Therefore, we do not expect the signal to be large at 0.2 $\micron$ for Earth-like planets. We note however that our calculations agree closely with simulations made using the \citet{robinson2016characterizing} model when including additional background gases (see Appendix for further details). 

The level of O$_2$ during the Proterozoic is highly uncertain \citep{reinhard2017false,olson2018earth}, with lower and upper constraints from various geochemical records and modeling efforts that vary by four orders of magnitude from 10$^{-5}$ to 10$^{-1}$ PAL \citep{pavlov2002mass, planavsky2014low, lyons2014rise, olson2018earth}. The lower limit on Proterozoic O$_2$ is inferred from the end of mass-independent fractionation of S isotopes in the wake of the ``Great Oxidation Event.'' The upper limit of 10$^{-1}$ PAL comes from the observation that the deep ocean remained anoxic throughout the Proterozoic, implying that surface environments were only mildly oxygenated. More recent work by \citet{planavsky2014low} provides an upper limit of 10$^{-3}$ PAL by leveraging the absence of Cr isotope fractionation in Proterozoic marine sediments. Although such geochemical data constrains the range of Proterozoic O$_2$ between 10$^{-5}$ and 10$^{-3}$ PAL, the low end of that range is difficult to explain in biogeochemical and photochemical models if oxygenic photosynthesis was occurring at or near modern rates \citep{ozaki2019sluggish}. A number of previous studies that considered the detectability of Proterozoic O$_2$ remained above an O$_2$ threshold of approximately 10$^{-3}$ PAL \citep{reinhard2017false, schwieterman2018exoplanet,gaudi2020habitable}. We considered Proterozoic O$_2$ levels as low as 10$^{-5}$ PAL to span the full range of estimates existing in the literature and permissible by existing geochemical data, and to allow us to consider every possible scenario in the search for biosignatures with HabEx and LUVOIR, but we note that the low estimate of 10$^{-5}$ PAL may be less likely than higher estimates.

In this work, we have based the definition of ``Earth-like'' on the premise that EECs would oxygenate roughly following Earth's oxygenation trajectory, and have used Earth history as a prior. However, the timing and timescale of oxygenation on Earth are not well understood, and it is possible that EECs would follow a different oxygenation trajectory. If Earth is rare and EECs never oxygenate past Archean levels, then as per our statistical definition of ``Earth-like'', EECs would not be statistically Earth-like. Similarly, if only a small fraction of EECs oxygenate past Archean levels, then statistically only a small fraction of EECs are Earth-like and f$_E$ is small. This is something we will be able to test with a mission that includes a large-aperture telescope, as whether or not we detect O$_3$, it will allow us to infer a maximum value of f$_E$. We have also explored other possible oxygenation scenarios for $\eta_{\earth}$ = 24\%: 1) Proterozoic O$_2$ levels are as high as modern levels, and 2) modern levels of O$_2$ are as low as those of the Proterozoic. In scenario 1, the distributions of detectable EECs with detectable O$_3$ are similar to those of Figure~\ref{fig:summary_nc}'s top panels (where Proterozoic O$_2$ is 0.1 PAL) for LUVOIR-A, LUVOIR-B, and HabEx/SS. That is because O$_3$ is detectable on every target EEC at every distance in the case where Proterozoic O$_2$ is 0.1 PAL, so assuming that Proterozoic levels are as high as modern levels will produce the same results. For HabEx/no-SS, the mission-level false negative scenario probability decreases from $1.8-36$\% to 1.5\%. In scenario 2, the distributions of detectable EECs with detectable O$_3$ are the same as those in Figure~\ref{fig:summary_nc} in most cases: for Proterozoic O$_2$ $\geq$10$^{-4}$ PAL for LUVOIR-A, for Proterozoic O$_2$ $\geq$10$^{-3}$ PAL for LUVOIR-B, and for Proterozoic O$_2$ $\geq$10$^{-4}$ PAL for HabEx/SS. That is because, again, for Proterozoic O$_2$ above those levels, O$_3$ is detectable on every target EEC at every distance. For lower Proterozoic O$_2$, there is still no mission-level false negative scenario probability for the 10$^{-5}$ PAL case for LUVOIR-A and for the 10$^{-4}$ PAL case for LUVOIR-B, but it increases to 1.3\% for 10$^{-5}$ PAL for LUVOIR-B and to 1.8\% for 10$^{-5}$ PAL for HabEx/SS. For HabEx/no-SS, the mission-level false negative scenario probability increases from $1.8-36$\% to $3.3-100$\%. Further oxygenation scenarios could be explored in future work.

Whether the origination of life is an extremely rare occurrence or common throughout the universe is a heavily debated topic. The frequency of life originating on habitable planets is highly uncertain \citep[e.g.,][]{sandberg2018dissolving}, and the fact that life originated on Earth does not constrain this frequency very much \citep{spiegel2012bayesian}. However, the large number of exoplanets that we may be able to soon characterize with future missions offers an opportunity to test whether the origin of life is common. If we observe a number of EECs and detect clearly biogenic O$_2$ or O$_3$ on at least one of them, the origination of life on habitable planets must be common. Conversely, if we don't detect O$_2$ or O$_3$ with a LUVOIR-A or LUVOIR-B-like instrument, we'll know that EECs are generally unlikely to be Earth-like. This could mean that either the origination of life is very rare, or that life rarely develops oxygenic photosynthesis. In this scenario of null-life detection, we may be able to improve our estimate of the probability of the origination of life using a Bayesian analysis similar to that of \citet{spiegel2012bayesian} and \citet{kipping2020objective}. Future work could look at what constraints we can put on the origination of life using a Bayesian analysis for different observation scenarios. 

\newpage

\section{Conclusions}
\label{sec:conclusions}

In this article, we considered whether observations of exoEarth candidates (EECs) with HabEx and LUVOIR may inform us on the fraction of EECs that are Earth-like (f$_E$) in that they develop Earth-like, O$_2$-producing life (oxygenic photosynthesis) and become oxygenated roughly following Earth's oxygenation history. To do that, we first considered the probability that HabEx and LUVOIR will detect O$_2$ and/or O$_3$ on EECs. Then, we determined whether a null detection, where we do not detect O$_2$ or O$_3$ on any EEC, would allow us to constrain f$_E$. We adopted a statistical approach to this problem. Instead of investigating false negatives on particular planets, we determined whether there might be mission-level false negatives for missions such as LUVOIR and HabEx based on all of the information that we can gain from all of the EECs that these missions can be expected to observe. We considered four different telescope designs: 15 m segmented on-axis LUVOIR-A, 8 m segmented off-axis LUVOIR-B, 4 m monolith off-axis HabEx with a starshade (``HabEx/SS'', where ``SS'' refers to HabEx's starshade) and 4 m monolith off-axis HabEx without a starshade (``HabEx/no-SS''). We also considered three different estimates for $\eta_{\earth}$: 24\%, following \citet{stark2019exoearth}, 5\%, following \citet{pascucci2019impact}, and 0.5\%, following \citet{neil2020joint}. In each case, we explore five different levels of Proterozoic O$_2$ (10$^{-1}$$-$10$^{-5}$ PAL), but we note that the lowest end of this range (10$^{-5}$ PAL) is less likely as it is difficult to explain if oxygenic photosynthesis was occurring at modern rates. Therefore these conclusions report our results for Proterozoic O$_2$ levels between 10$^{-1}$$-$10$^{-4}$ PAL. The main conclusions of this article are:

\begin{enumerate}

\item First, we considered the possibility of not detecting any EECs. The probability of that occurring depends strongly on both $\eta_{\earth}$ and the mirror diameter of the instrument. For the cases we considered, these probabilities are:
\begin{itemize}
\item For $\eta_{\earth}$ = 24\%: 0\% for all missions.
\item For $\eta_{\earth}$ = 5\%: 11\% for HabEx, 0\% for LUVOIR-A and LUVOIR-B.
\item For $\eta_{\earth}$ = 0.5\%: 27\%, 50\%, and 81\%, for LUVOIR-A, LUVOIR-B, and HabEx.
\end{itemize}

\item Second, we considered the possibility of mission-level false negative scenarios where we do not detect O$_2$ or O$_3$ on any of the EECs we observe, even if they were all Earth-like (f$_E$ = 1). We find that the main factor determining whether such mission-level false negative scenarios exist is the yield of detectable EECs, which mainly depends on the value of $\eta_{\earth}$ and on the instrument's mirror diameter. 
We found that the level of Proterozoic O$_2$ we assume does not matter much for these missions, as long as it is above 10$^{-4}$ PAL, as in that case O$_3$ is detectable on all target EECs with LUVOIR-A, LUVOIR-B, HabEx/SS. The probabilities of a mission-level false negative scenario are as follows:
\begin{itemize}
\item For $\eta_{\earth}$ = 24\%: 0\% for LUVOIR-A and LUVOIR-B; 0.5\% for HabEx/SS; 2$-$36\% for HabEx/no-SS.
\item For $\eta_{\earth}$ = 5\%: 0.1\% for LUVOIR-A; 3-4\% for LUVOIR-B; 33\% for HabEx/SS; 44$-$81\% for HabEx/no-SS.
\item For $\eta_{\earth}$ = 0.5\%: 52\% for LUVOIR-A; 70$-$72\% for LUVOIR-B; 90\% for HabEx/SS; 91-98\% for HabEx/no-SS.
\end{itemize}
 
\item Finally, we considered whether we could constrain the fraction of EECs that are Earth-like (f$_E$) even if we do not detect O$_2$ or O$_3$ on any EEC. The extent to which we may be able to constrain f$_E$ in the case of a null detection depends on the mission and relies primarily on a sufficiently large number of detectable EECs as well as on the ability to detect low levels of O$_3$. 
\begin{itemize}
\item For $\eta_{\earth}$ = 24\%, a null detection with LUVOIR-A, LUVOIR-B, or HabEx/SS will allow us to constrain the fraction of Earth-like EECs: $f_E \leq 0.094$ with LUVOIR-A , $f_E \leq 0.18$ with LUVOIR-B, and $f_E \leq 0.56$ with HabEx/SS, all with 95\% confidence and for Proterozoic O$_2$ levels $\geq$ 10$^{-4}$ PAL. These constraints also imply that for f$_E$ greater than these upper limits, we have a 95\% chance of detecting O$_3$ on at least 1 EEC, and are therefore likely to do so with LUVOIR-A for f$_E$ $\geq$ 0.094, with LUVOIR-B for f$_E$ $\geq$ 0.18, and with HabEx for f$_E$ $\geq$ 0.56.
\item For $\eta_{\earth}$ = 5\%, a null detection with LUVOIR-A or LUVOIR-B would similarly allow us to constrain $f_E \leq 0.45$ with LUVOIR-A, and $f_E \leq 0.85$ with LUVOIR-B, with 95\% confidence and for Proterozoic O$_2$ levels $\geq$ 10$^{-4}$ PAL. HabEx/SS and HabEx/no-SS on the other hand will not allow us to constrain $f_E$ due to the existence of mission-level false negative scenarios even for f$_E$ = 1. This also means that we may not detect O$_3$ on any EECs with HabEx/SS and HabEx/no-SS, while we are likely to do so with LUVOIR-A for $f_E \geq 0.45$, and with LUVOIR-B for $f_E \geq 0.85$.
\item For $\eta_{\earth}$ = 0.5\%, none of the missions would allow us to constrain $f_E$ in the event of a null detection, as all have mission-level false negative scenarios with a probability greater than 5\%. Therefore, despite the fact that missions with larger aperture mirrors are more robust to uncertainties in $\eta_{\earth}$, all missions are vulnerable to inconclusive null detections if $\eta_{\earth}$ is as low as 0.5\%. 

\end{itemize}
\end{enumerate}

\section{Acknowledgements}
\label{sec:acknowledgements}

We thank our anonymous reviewer for a constructive and insightful review that helped us better this manuscript. We thanks Chris C. Stark for sharing and guiding us through his target list database for HabEx and LUVOIR. We also thank him, as well as Giada Arney, for their help in verifying assumptions made for the LUVOIR mission design. We thank Jacob L. Bean for insightful discussions on the detectability of O$_2$ and O$_3$. This work was supported by the NASA Astrobiology Program Grant Number 80NSSC18K0829 and benefited from participation in the NASA Nexus for Exoplanet Systems Science research coordination network. S.L.O. acknowledges support from the T.C. Chamberlin Postdoctoral Fellowship in the Department of the Geophysical Sciences at the University of Chicago. SLO additionally acknowledges support from the NASA Habitable Worlds Program. T.D.K. acknowledges funding from the 51 Pegasi b Fellowship in Planetary Astronomy sponsored by the Heising-Simons Foundation. P.P. acknowledges support from the James S. McDonnell Foundation. 

\newpage

\appendix

Computing the integration times required for a detection of O$_2$ or O$_3$ requires two simulation components: a planetary spectrum model and a noise model. Molecular signatures for each atmospheric species of interest were computed by taking the difference between the continuum (all atmospheric species except the species of interest) and absorbance (all atmospheric species) spectra. The noise and SNR calculations involve the characterization of the different factors impacting the sensitivity. 
The signal refers to the planetary photons that are successfully counted by the instrument, after considering the telescope, coronagraph/starshade, optics and detector efficiencies. The noise is a combination of effects that includes the Poisson noise introduced by the planetary photons, noise from residual photons from the star, noise from background sources (e.g., exozodi and local zodiacal fluxes), and the intrinsic noise introduced by the detector. 

In order to further expand on these components, the appendix is therefore structured as follows: 1) we show geometric albedo spectra used to calculate the SNR of O$_2$ and O$_3$ detections with PSG, 2) we present integration times required for a 5-$\sigma$ detection of O$_2$ and O$_3$ at 10, 15, and 20 pc calculated using PSG, 3) we compare our calculated integration times to those calculated using \citet{robinson2016characterizing}'s model, 4) we further detail the assumptions made in simulating observations with LUVOIR and HabEx using PSG.
\bigskip

\section{Geometric Albedo Spectra}
\label{sec:geom}

\begin{figure*}[h!]
\begin{center}
  \includegraphics[width=1.\textwidth]{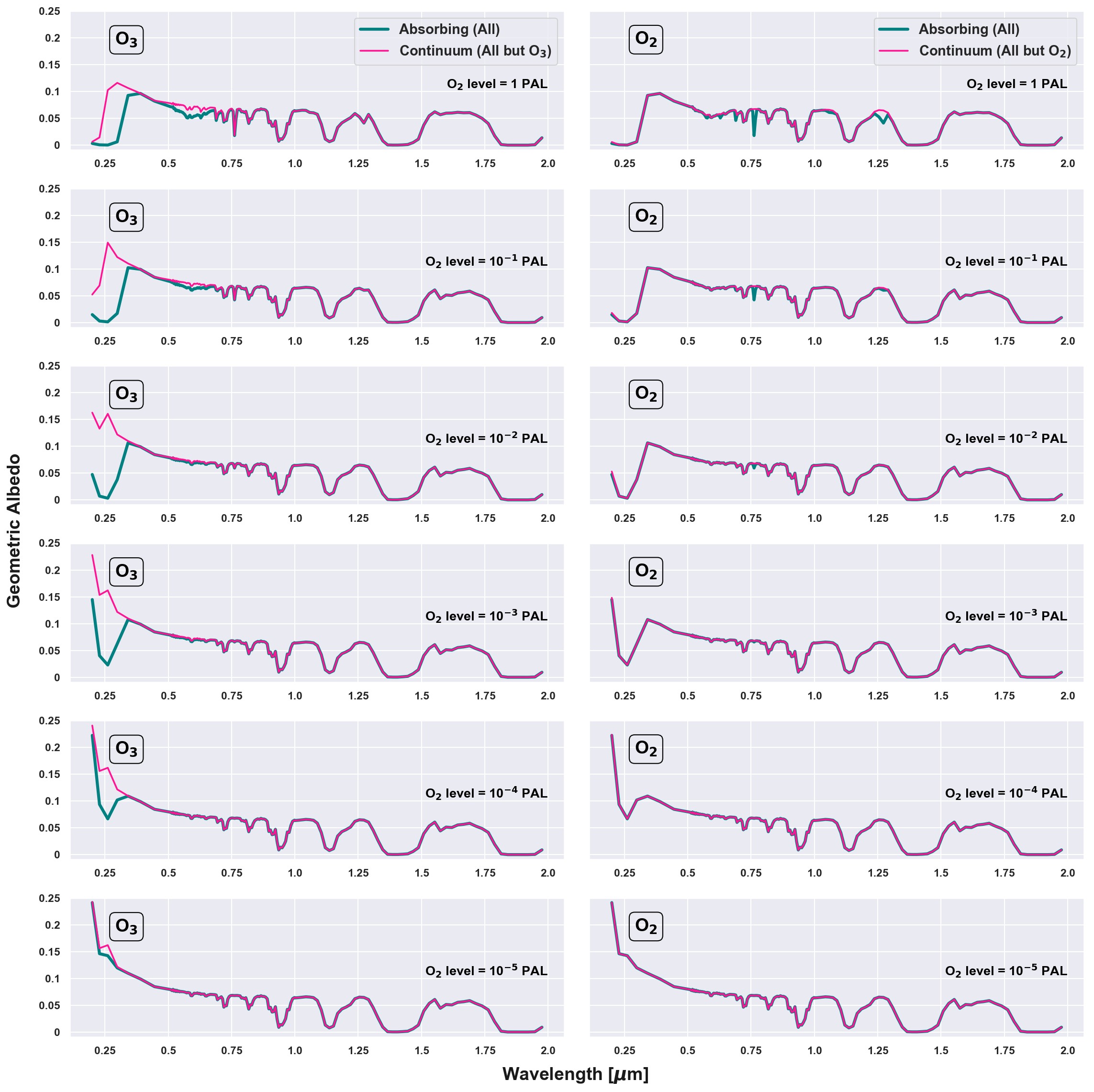}
  \caption{\textbf{Continuum and absorption spectra for O$_3$ (left) and O$_2$ (right) detection for an Earth-like planet orbiting at 1 AU around a Sun-like star at 5 pc with an exposure time of 1000 hours.} We consider six different O$_2$ levels from 10$^{-5}$ to 1 PAL. Background gases include gases such as N$_2$, H$_2$O, CH$_4$, and CO$_2$.}
  \label{fig:geomalb}
\end{center}
\end{figure*}

As described in Section~\ref{sec:snrcalc}, we calculate the SNR of O$_2$ and O$_3$ detections by first simulating two spectra: one absorbing spectrum with all atmospheric species, and one continuum spectrum with all atmospheric species except the chosen absorber (either O$_2$ or O$_3$). In Figure~\ref{fig:geomalb}, we show both spectra for O$_3$ (left) and O$_2$ (right) at the six O$_2$ levels considered in this work. 

O$_3$'s strong feature at $\sim$0.25 $\mu$m allows for it to be detected even at very low levels. On the other hand, O$_2$'s strongest feature at 0.76 $\mu$m becomes difficult to detect at Proterozoic O$_2$ levels. We note here that we added the collision-induced O$_2$-O$_2$ absorption bands in the UV (Wulf bands) between 0.24-0.3 $\mu$m as well as the Herzberg O$_2$ continuum bands \citep{fally2000fourier}, and the Herzberg O$_2$ band system  \citep{jenouvrier1999fourier,merienne2000fourier,merienne2001improved}, neither of which are included in the HITRAN database.

\newpage 

\section{Integration times at 10, 15, and 20 pc}
\label{sec:times}

Here we present the integration times required for a 5-$\sigma$ detection of O$_2$ and O$_3$ at 10, 15, and 20 pc calculated using PSG \citep{villanueva2018planetary} in Tables~\ref{tab:SNRs10},~\ref{tab:SNRs15}, and~\ref{tab:SNRs20} for LUVOIR-A, LUVOIR-B, HabEx/SS, and HabEx/no-SS (see Section~\ref{sec:snrcalc} for details and for calculations at 5 pc).

\begin{center}
\begin{table*}[h!]
\centering
  \caption{\textbf{Integration times [hrs] with LUVOIR-A (15 m), LUVOIR-B (8 m), HabEx/SS (4 m), and HabEx/no-SS (4 m) to yield a 5-$\sigma$ detection of O$_2$ and O$_3$ for an Earth-like planet without clouds at 10 pc for six different O$_2$ levels.}}
\label{tab:SNRs10}
\centering
\begin{tabular}{l l l l l}
    &  15 m LUVOIR-A & 8 m LUVOIR-B & 4 m HabEx/SS & 4 m HabEx/no-SS \\
\hline
O$_2$ = 1 PAL  &  O$_2$: 5.93 hrs &  O$_2$: 47.7 hrs & O$_2$: 182.3 hrs & O$_2$: 1121.5 hrs\\ 
   &  O$_3$: 1.04 hr &  O$_3$: 4.6 hr & O$_3$: 3.33 hr & O$_3$: 332.8 hrs \\ 
O$_2$ = 10$^{-1}$ PAL  &  O$_2$: 30.1 hrs &  O$_2$: 241.1 hrs & O$_2$: 877.8 hrs & O$_2$: 5376.6 hrs\\ 
   &  O$_3$: 1.71 hr &  O$_3$: 6.68 hrs & O$_3$: 2.92 hr & O$_3$: 1217.6 hrs \\ 
O$_2$ = 10$^{-2}$ PAL  & O$_2$: 288.5 hrs &  O$_2$: 2310.8 hrs & O$_2$: 7233.8 hrs & O$_2$: 5.1 $\times 10^4$ hrs\\ 
  &  O$_3$: 2.94 hrs &  O$_3$: 10.8 hrs & O$_3$: 3.12 hr & O$_3$: 1.1 $\times 10^4$ hrs \\ 
O$_2$ = 10$^{-3}$ PAL  & O$_2$: 4603.8 hrs &  O$_2$: 3.7 $\times 10^4$ hrs & O$_2$: 7.5 $\times 10^4$ hrs & O$_2$: 8.1 $\times 10^5$ hrs \\ 
   & O$_3$: 5.24 hrs &  O$_3$: 19.31 hrs & O$_3$: 4.73 hr & O$_3$: 2.8 $\times 10^4$ hrs \\ 
O$_2$ = 10$^{-4}$ PAL  & O$_2$: 2.1 $\times 10^5$ hrs &  O$_2$: 1.7 $\times 10^6$ hrs & O$_2$: 1.4 $\times 10^6$ hrs & O$_2$: 3.8 $\times 10^7$ hrs \\ 
   & O$_3$: 23.53 hrs &  O$_3$: 89.33 hrs & O$_3$: 14.73 hrs & O$_3$: 1.8 $\times 10^6$ hrs \\ 
O$_2$ = 10$^{-5}$ PAL  & O$_2$: 1.5 $\times 10^8$ hrs &  O$_2$: 1.9 $\times 10^7$ hrs & O$_2$: 7.9 $\times 10^7$ hrs & O$_2$: 3.3 $\times 10^9$ hrs\\ 
   & O$_3$: 740.5 hrs &  O$_3$: 2884.0 hrs & O$_3$: 387.9 hrs & O$_3$: 3.8 $\times 10^8$ hrs \\ 
\end{tabular}
\end{table*}
\end{center}

\begin{center}
\begin{table*}[h!]
\centering
  \caption{\textbf{Integration times [hrs] with LUVOIR-A (15 m), LUVOIR-B (8 m), HabEx/SS (4 m), and HabEx/no-SS (4 m) to yield a 5-$\sigma$ detection of O$_2$ and O$_3$ for an Earth-like planet without clouds at 15 pc for six different O$_2$ levels.}}
\label{tab:SNRs15}
\centering
\begin{tabular}{l l l l l}
    &  15 m LUVOIR-A & 8 m LUVOIR-B & 4 m HabEx/SS & 4 m HabEx/no-SS \\
\hline
O$_2$ = 1 PAL  &  O$_2$: 30.56 hrs &  O$_2$: 320.0 hrs & O$_2$: 951.7 hrs & O$_2$: 1.6 $\times 10^4$ hrs\\ 
   &  O$_3$: 3.34 hr &  O$_3$: 20.7 hr & O$_3$: 13.2 hr & O$_3$: 2963.5 hrs \\ 
O$_2$ = 10$^{-1}$ PAL  &  O$_2$: 154.8 hrs &  O$_2$: 1614.1 hrs & O$_2$: 4411.1 hrs & O$_2$: 8.0 $\times 10^4$ hrs\\ 
   &  O$_3$: 5.27 hr &  O$_3$: 28.31 hrs & O$_3$: 11.27 hr & O$_3$: 1.1 $\times 10^4$ hrs \\ 
O$_2$ = 10$^{-2}$ PAL  & O$_2$: 1485.9 hrs &  O$_2$: 1.5 $\times 10^4$ hrs & O$_2$: 3.6 $\times 10^4$ hrs & O$_2$: 7.6 $\times 10^5$ hrs\\ 
  &  O$_3$: 8.93 hrs &  O$_3$: 44.8 hrs & O$_3$: 11.88 hrs & O$_3$: 9.5 $\times 10^4$ hrs \\ 
O$_2$ = 10$^{-3}$ PAL  & O$_2$: 2.4 $\times 10^4$ hrs &  O$_2$: 2.5 $\times 10^5$ hrs & O$_2$: 3.5 $\times 10^5$ hrs & O$_2$: 1.2 $\times 10^7$ hrs \\ 
   & O$_3$: 16.2 hrs &  O$_3$: 81.2 hrs & O$_3$: 17.85 hr & O$_3$: 2.5 $\times 10^5$ hrs \\ 
O$_2$ = 10$^{-4}$ PAL  & O$_2$: 1.1 $\times 10^6$ hrs &  O$_2$: 1.1 $\times 10^7$ hrs & O$_2$: 5.8 $\times 10^6$ hrs & O$_2$: 5.7 $\times 10^8$ hrs \\ 
   & O$_3$: 77.1 hrs &  O$_3$: 394.7 hrs & O$_3$: 55.1 hrs & O$_3$: 1.7 $\times 10^7$ hrs \\ 
O$_2$ = 10$^{-5}$ PAL  & O$_2$: 9.8 $\times 10^7$ hrs &  O$_2$: 1.0 $\times 10^9$ hrs & O$_2$: 3.3 $\times 10^8$ hrs & O$_2$: 5.0 $\times 10^{10}$ hrs\\ 
   & O$_3$: 2532.2 hrs &  O$_3$: 1.3 $\times 10^4$ hrs & O$_3$: 1442.8 hrs & O$_3$: 3.4 $\times 10^9$ hrs \\ 
\end{tabular}
\end{table*}
\end{center}

\begin{center}
\begin{table*}[h!]
\centering
  \caption{\textbf{Integration times [hrs] with LUVOIR-A (15 m), LUVOIR-B (8 m), HabEx/SS (4 m), and HabEx/no-SS (4 m) to yield a 5-$\sigma$ detection of O$_2$ and O$_3$ for an Earth-like planet without clouds at 20 pc for six different O$_2$ levels.}}
\label{tab:SNRs20}
\centering
\begin{tabular}{l l l l l}
    &  15 m LUVOIR-A & 8 m LUVOIR-B & 4 m HabEx/SS & 4 m HabEx/no-SS \\
\hline
O$_2$ = 1 PAL  &  O$_2$: 95.84 hrs &  O$_2$: 1564.0 hrs & O$_2$: 1.4 $\times 10^4$ hrs & O$_2$: 1.6 $\times 10^5$ hrs\\ 
   &  O$_3$: 9.12 hr &  O$_3$: 69.71 hr & O$_3$: 38.67 hr & O$_3$: 2.4 $\times 10^4$ hrs \\ 
O$_2$ = 10$^{-1}$ PAL  &  O$_2$: 465.8 hrs &  O$_2$: 8033.6 hrs & O$_2$: 6.4 $\times 10^4$ hrs & O$_2$: 8.9 $\times 10^5$ hrs\\ 
   &  O$_3$: 13.2 hr &  O$_3$: 91.1 hrs & O$_3$: 31.9 hr & O$_3$: 8.9 $\times 10^4$ hrs \\ 
O$_2$ = 10$^{-2}$ PAL  & O$_2$: 4441.8 hrs &  O$_2$: 7.7 $\times 10^4$ hrs & O$_2$: 3.4 $\times 10^5$ hrs & O$_2$: 8.6 $\times 10^6$ hrs\\ 
  &  O$_3$: 21.6 hrs &  O$_3$: 142.1 hrs & O$_3$: 33.23 hr & O$_3$: 7.8 $\times 10^5$ hrs \\ 
O$_2$ = 10$^{-3}$ PAL  & O$_2$: 7.1 $\times 10^4$ hrs &  O$_2$: 1.2 $\times 10^6$ hrs & O$_2$: 1.7 $\times 10^6$ hrs & O$_2$: 1.4 $\times 10^8$ hrs \\ 
   & O$_3$: 39.4 hrs &  O$_3$: 258.9 hrs & O$_3$: 49.7 hr & O$_3$: 2.1 $\times 10^6$ hrs \\ 
O$_2$ = 10$^{-4}$ PAL  & O$_2$: 3.3 $\times 10^6$ hrs &  O$_2$: 5.8 $\times 10^7$ hrs & O$_2$: 1.9 $\times 10^7$ hrs & O$_2$: 6.5 $\times 10^9$ hrs \\ 
   & O$_3$: 195.4 hrs &  O$_3$: 1292.4 hrs & O$_3$: 152.3 hrs & O$_3$: 1.4 $\times 10^8$ hrs \\ 
O$_2$ = 10$^{-5}$ PAL  & O$_2$: 2.9 $\times 10^8$ hrs &  O$_2$: 5.1 $\times 10^9$ hrs & O$_2$: 1.0 $\times 10^9$ hrs & O$_2$: 5.7 $\times 10^{11}$ hrs\\ 
   & O$_3$: 6625.7 hrs &  O$_3$: 4.4 $\times 10^4$ hrs & O$_3$: 3983.3 hrs & O$_3$: 2.9 $\times 10^{10}$ hrs \\ 
\end{tabular}
\end{table*}
\end{center}

\newpage 

\section{Comparison to Robinson et al. (2016) model}
\label{sec:comp}

\begin{figure*}[h!]
\begin{center}
  \includegraphics[width=1.\textwidth]{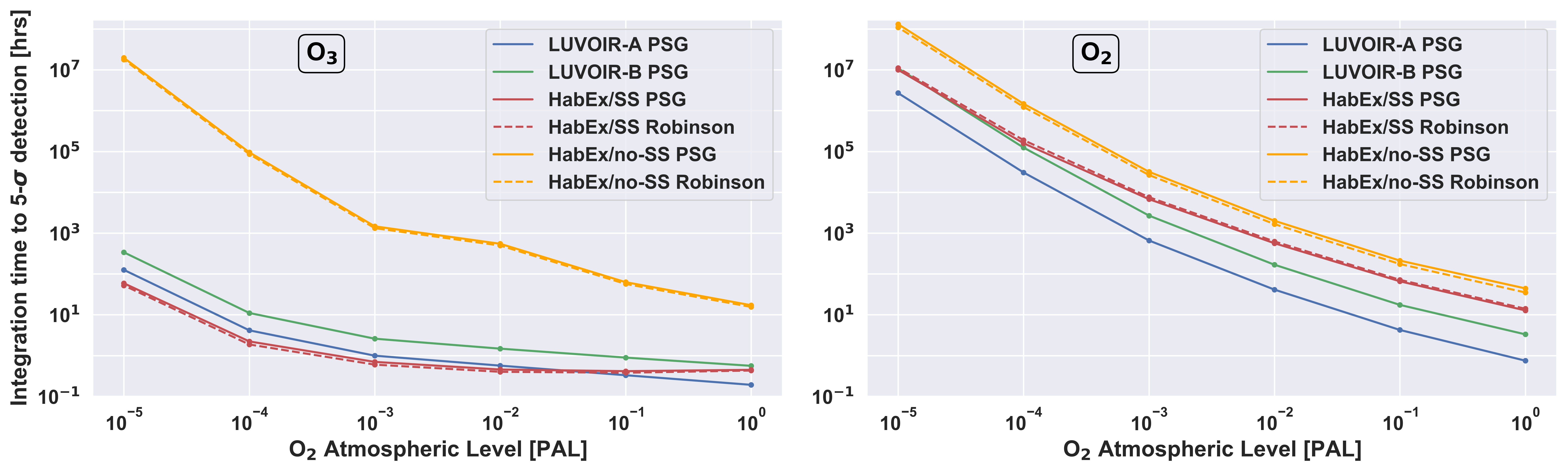}
  \caption{\textbf{Integration times necessary to detect O$_3$ (left) and O$_2$ (right) at 5-$\sigma$ as a function of O$_2$ atmospheric level, for an Earth-like planet orbiting at 1 AU around a Sun-like star at 5 pc.} Solid lines: PSG calculations, Dashed lines: \citet{robinson2016characterizing}'s model. Blue: LUVOIR-A, Green: LUVOIR-B, Red: HabEx/SS, Orange: HabEx/no-SS.}
  \label{fig:tycompare}
\end{center}
\end{figure*}

In Figure~\ref{fig:tycompare} we compare the integration times we calculated using PSG to those calculated using the \citet{robinson2016characterizing} model for an Earth-like planet at 5 pc. In both cases, we calculate the integration times following the method outlined in Section~\ref{sec:snrcalc}. We note here that \citet{robinson2016characterizing}'s model was updated for these calculations according to the latest figures reported in the Final Reports \citep{luvoir2019luvoir,gaudi2020habitable} and we include Earth-like background gases such as O$_2$, O$_3$, N$_2$, H$_2$O, CO$_2$, and CH$_4$ in the atmosphere. Therefore, the \citet{robinson2016characterizing} model calculations may not match those reported in Figure 3.3-7 of the HabEx report \citep{gaudi2020habitable}.

The integration times calculated using each model agree with each other very closely. This Figure also exhibits the fact that HabEx/SS performs very well in the UV thanks to its starshade and despite its small size. Because of that, it outperforms LUVOIR-A and LUVOIR-B at low O$_3$ concentrations. 

\newpage
\section{Simulation parameters}
\label{sec:corona}

\begin{figure*}[h!]
\begin{center}
  \includegraphics[width=0.8\textwidth]{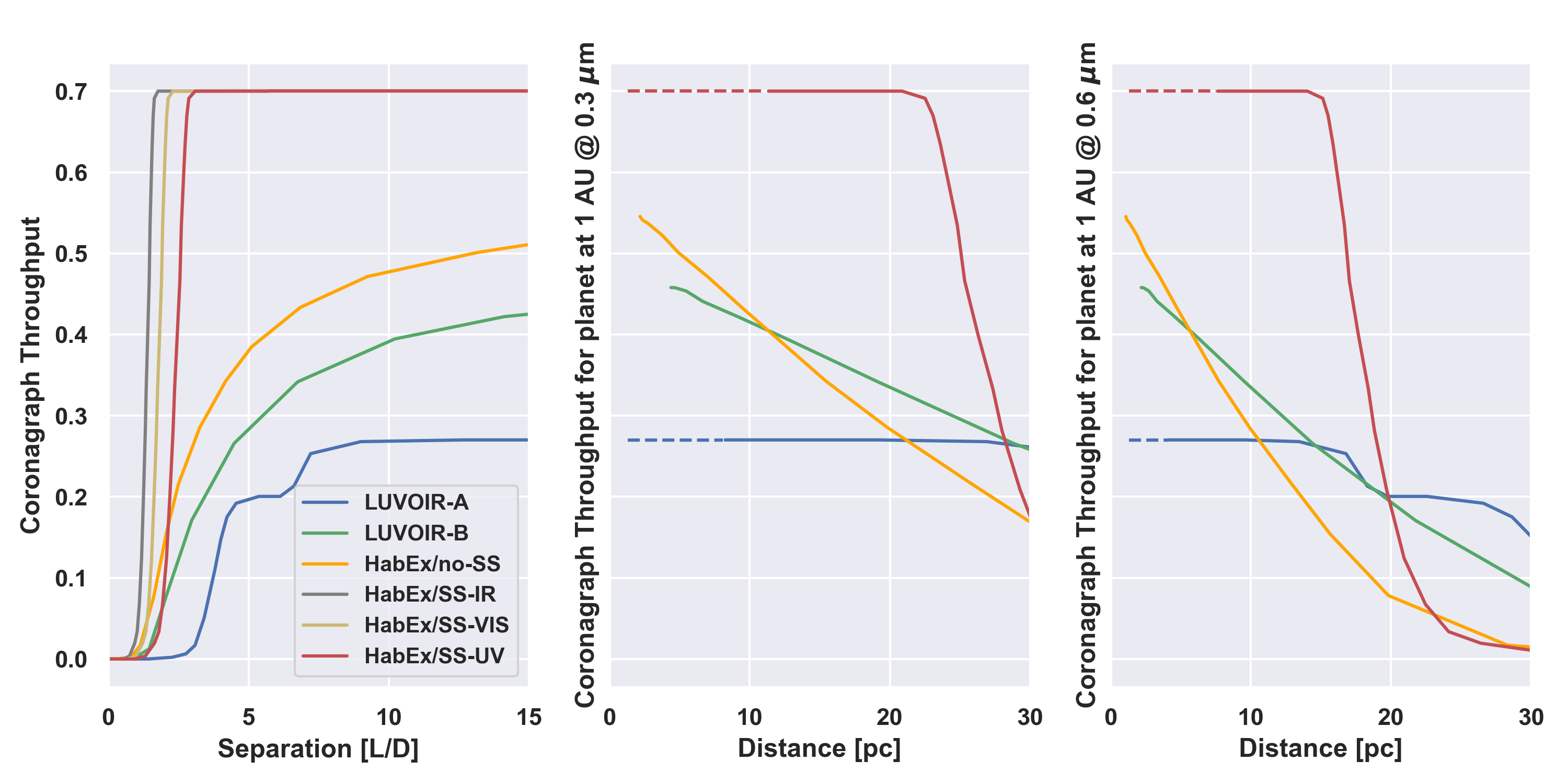}
  \caption{\textbf{Coronagraph throughput of LUVOIR-A, LUVOIR-B, HabEx/SS, and HabEx/no-SS as a function of separation (left). Coronagraph throughput at 0.3 $\mu$m (center) and at 0.6 $\mu$m (right) as a function of distance.} Following \citet{stark2019exoearth}, LUVOIR-A corresponds to the APLC coronagraph with a segmented on-axis OTA, LUVOIR-B corresponds to the DMVC coronagraph with a segmented on-axis OTA, HabEx/no-SS corresponds to a vortex coronagraph with an off-axis monolithic OTA. The HabEx/SS throughput was taken from Figure 6.4-3 in the HabEx final report and integrates a 70$\%$ loss due to PSF losses. }
  \label{fig:corona}
\end{center}
\end{figure*}

The planetary photons being measured by the detector go through a series of optical systems, that each can be assumed to have a specific efficiency or throughput. Background sources also go through the same optical path as the planetary photons. We define the end-to-end throughput for the planetary fluxes as: $T_{total} = T_{Tele} \times T_{coronagraph} \times T_{opt} \times T_{read} \times T_{QE}$, where $T_{Tele}$ accounts for light lost due to contamination and inefficiencies in the main collecting area, $T_{coronagraph}$ is the coronagraphic throughput at this planet-star separation, $T_{opt}$ is the optical throughput (the transmissivity of all optics), $T_{QE}$ is the raw quantum efficiency (QE) of the detector, and $T_{read}$ is the read-out efficiencies. A summary and representative value for each these parameters can be found in Table~\ref{tab:params} and we show the coronagraph throughput as a function of separation in Figure~\ref{fig:corona}.

For $T_{Tele}$, we adopt 0.95 for all wavelengths, on par with the particulate coverage fraction for JWST’s mirrors. EMCCD detectors are expected to have $T_{read}$ near 0.75 \citep{stark2019exoearth}, while for NIR and other detectors, read-out inefficiencies and bad-pixels may account to a similar value, and we simply adopt $T_{read}$=0.75 across all detectors as a conservative estimate. The reported quantum efficiency of the different detectors ranges from 0.6 to 0.9 (LUVOIR and HabEx reports \citep{luvoir2019luvoir,gaudi2020habitable}), yet technological improvements in several of these detectors could be expected in the near future, and we adopt a general $T_{QE}$=0.9 for all detectors, bands, and for both observatories.

\begin{figure*}[h!]
\begin{center}
  \includegraphics[width=0.5\textwidth]{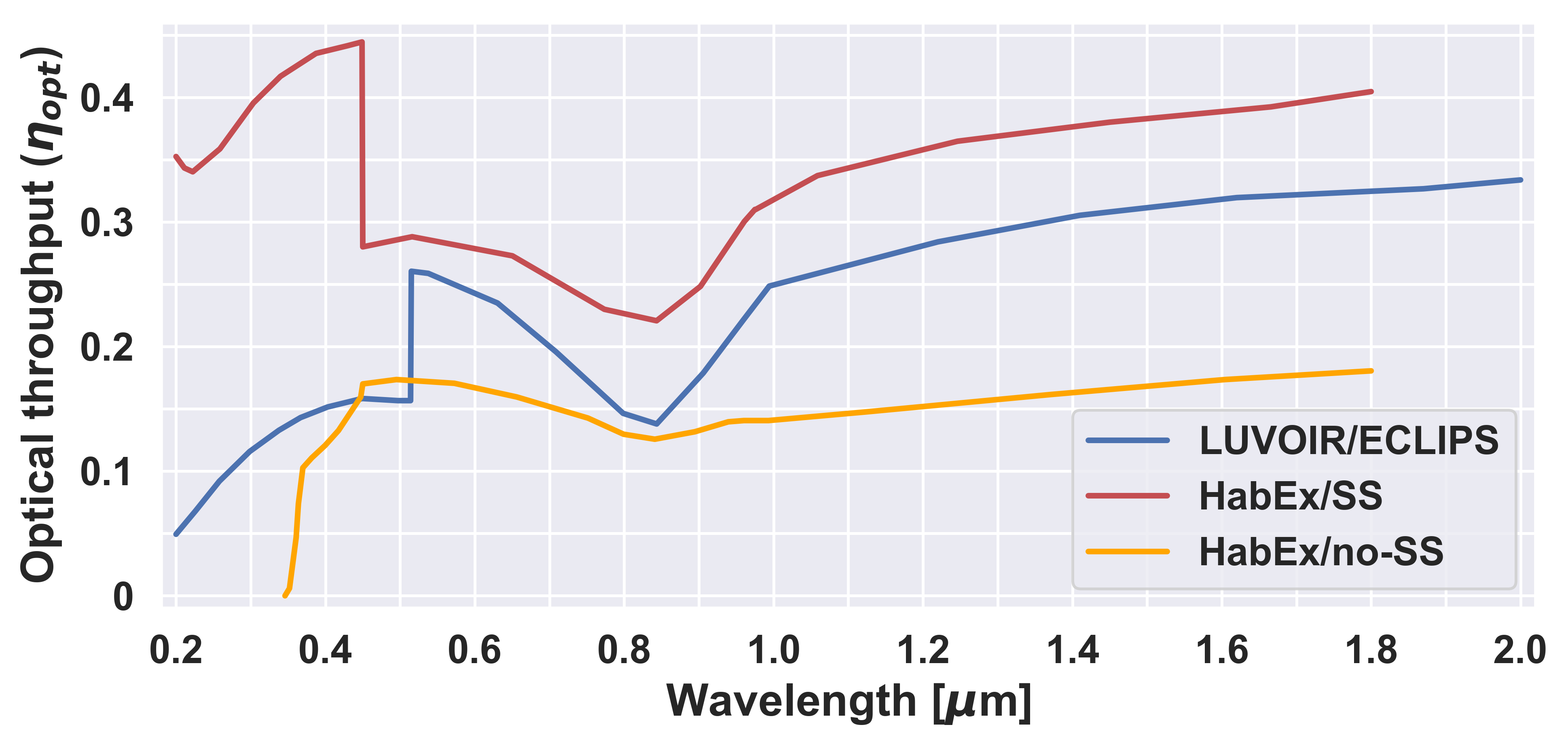}
  \caption{\textbf{Optical throughput of LUVOIR, HabEx/SS, and HabEx/no-SS as a function of wavelength.} Optical efficiencies for HabEx were taken from the final report \citep{gaudi2020habitable}, considering the IFS mode for the visible and infrared channels. For LUVOIR, the optical efficiencies were captured from the final report \citep{luvoir2019luvoir} and from Figure 4 (IFS mode) of \citet{stark2019exoearth}.}
  \label{fig:topt}
\end{center}
\end{figure*}

Optical efficiencies ($T_{opt}$) for HabEx were taken from the HabEx report \citep{gaudi2020habitable}, specifically from Figure 6.4-10 for HabEx/SS and from Figure 6.3-6 for HabEx/no-SS, considering the IFS mode for the visible and infrared channels. For LUVOIR, the optical efficiencies were taken from the final report \citep{luvoir2019luvoir} and from Figure 4 (IFS mode) of \citet{stark2019exoearth}. These are shown in Figure~\ref{fig:topt}.

\bibliography{Bibs}

\end{document}